\RequirePackage{amsmath}
\documentclass[12pt]{iopart}
\usepackage{epsfig}
\usepackage{amssymb}
\usepackage{lineno}
\usepackage{graphicx}
\usepackage{xcolor}
\usepackage{mathtools}
\usepackage{iopams}
\begin{document}

\title[]{Microscopic study of the Shell Structure evolution in isotopes of light to middle mass range Nuclides}

\author{Virender Thakur$^{1,2}$, Pankaj Kumar$^{1}$, Suman Thakur$^{1}$, Smriti Thakur$^{1}$, Vikesh Kumar$^{1}$ \& Shashi K Dhiman$^{1,3}$}

\address{$^{1}$Department of Physics Himachal Pradesh University, Summer-Hill, Shimla-171005, INDIA}
\eads{\mailto{$^{2}$virenthakur2154@gmail.com}, \mailto{$^{3}$shashi.dhiman@gmail.com}}
\paper{}
\vspace{10pt}
\begin{indented}
\item[]January 2020
\end{indented}

\begin{abstract}
The shell structure of even-even isotopes in Si, S, Ar and Ca has been analysed. The theoretical calculations of
shell closure parameter $D_{n} (N)$ and the differential variation of the two-neutron separation energy $dS_{2n} (Z, N )$ are carried out within the framework of Hartree-Fock-Bogoliubov theory. Calculations are carried out for different skyrme forces and their sensitivity has been tested. Same nuclides are studied  by employing the relativistic mean field aproach based on meson exchange and point coupling models. Theoretically calculated estimates are in good agreement with the recently available experimental data which fortifies signature of shell closure at N = 14 and
20 in case of Si, N = 14, 20 and 28 in S and N=20 and 28 in  Ar and Ca isotopes.
\end{abstract}
\vspace{2pc}
\vspace{2pc}

\submitto{\jpg}
\maketitle
%
\section{Introduction}
\label{intro}
 In nuclear physics, understanding the structure of the atomic nucleus is one of the key challenges. Certain configurations of nucleons (magic numbers) in nucleus are observed to be more stable (magic nuclei). Investigating the evolution of these magic nuclei with nucleon number towards the dripline is an fascinating  area of research in  nuclear physics \cite{Holt2012}.  The study of those nuclei which are lying far from the line of $\beta$-stability play a crucial role in the understanding of nuclear physics. The production of the new isotopes \cite{Bhattacharyya2008,Stroberg2014}  in recent years has revived  the interest in nuclear structure models. This provoked the
development of radioactive ion beam (RIB) facilities and highly sensitive detectors to bring new discoveries \cite{Lunderberg2016,Gade2009}. Till today, very
little information is available about those nuclei which are lying near driplines (exotic or halo nuclei).\\
Even though the number of undiscovered bound nuclei is very large but we are able to
make single big steps by studying a few of specific nuclei. These nuclei act as milestones
in setting  new effects that arise in extreme conditions of isospin asymmetry. We have chosen Si, S, Ca and Ar nuclei for our interest as lot of research has been done on these nuclei recently \cite{Bhattacharyya2008,Stroberg2014,Thakur2019study}. The microscopic structure of these nuclei is
of particular interest for the field of  astrophysics: the neutron-rich N$\simeq$28 nuclei play an important
role in the nucleosynthesis of the heavy Ca-Ti-Cr isotopes. As these nuclides also become
experimentally accessible, they can provide a testing ground for studying exotic
nuclei. In light to middle  mass nuclei, the disappearance
of the traditional magic numbers and the appearance of new numbers far from the
stability line is one of the main issues in the nuclear structure physics of exotic nuclei. In ref. \cite{Ebata2015,Bastin2007}, the weakening of the shell closure for N = 28 has been reported  by investigating
the experimental $2^{+}$ excitation energies of Si isotopes. Similarily, the conventional
shell structure picture of magic numbers of nucleons in stable nuclides is no longer
applicable to the many exotic nuclei both in theoretical and experimental extractions  as reported in ref. \cite{Bastin2007,Campbell2006,Takeuchi2012,Stroberg2014} for $^{42}Si$ and in ref. \cite{Force2010,Kimura2013,Sarazin2000,Gaudefroy2009} for $^{42}Si$. In the chain of
Ca isotopes, experimental studies strongly mark N = 32 as a new magic number
due to the high energy of the first $2^{+}$ state in its nucleus \cite{Huck1985}\\ To study the physics of shell structure evolution in  exotic neutron-rich and magic-nuclei of Si, S, Ar and Ca isotopes  near drip lines, we
performed calculations based on the self-consistent mean-field theories: Hartree-Fock-Bogoliubov
(HFB) with the Skyrme effective interactions \cite{Perez2017,Skyrme1959}. Also, the three-dimensional relativistic Hartree-Bogoliubov (RHB) model with density dependent meson exchange (DD-ME2)
and point coupling (DD-PC1 and DD-PCX \cite{Yuksel2019}) effective interactions, are employed with separable pairing interac-
tion \cite{Nikvsic2014,Nikvsic2010,Tian2009}. The physical observables of interest are nuclear binding en-
ergies, nuclear shell closure parameters $D_{n}$(N). Results of  two-neutron separation energies
$S_{2n}$(Z,N) and differential variation in two-neutron separation energies $dS_{2n}$ (Z,N) are also presented.\\
The paper is organized as follows. A general overview of  HFB model with  effective skyrme interaction and the RHB model for density dependent meson exchange (DD-ME)
and point coupling (DD-PC) effective interactions is presented in
Section 2. In Section 3, the results of our theoretical calculations are presented and compared with the recently available experimental data. The results of the present work are summarized in Section 4.
\section{Theoretical Framework}
The presented work has been done by using the models based on   Hartree-Fock-Bogoliubov Theory \cite{Ring1980,Stoitsov2013,Perez2017} and  Relativistic Hartee Bogoliubov (RHB) Theory \cite{Nikvsic2011}. A brief discussion of these models is given below.\\
\subsection{HFB Theory}
This theory is a combination of Hartree-Fock (HF) and BCS theory. In this theory, the pairing field and the mean field are given equal footings. Details of the theory can be extracted from our previous work \cite{Thakur2019} based on the zero range Skyrme effective interactions \cite{Skyrme1959} used in the mean field part. More details can be found in ref. \cite{Ring1980,Stoitsov2013,Perez2017}. Out of different types of Skyrme forces like SLY5 \cite{Chabanat1997}, UNEDF0 \cite{Kortelainen2010}, UNEDF1 \cite{Kortelainen2012}, SKP \cite{Bartel1982}, and SKM* \cite{Dobaczewski1984} etc. that exists in the literature, we have adopted SLY5, UNEDF0 and SKM* parameterizations because of their best efficiency in reproducing the ground state properties of the nuclides of our interest. For the Skyrme forces, The HFB Energy has the form of local energy density functional \cite{Stoitsov2005,Stoitsov2013, Perez2017}. \\
In the particle-particle (pairing) channel, pairing interaction is including the density dependent delta interaction \cite{Chasman1976,Terasaki1997} of the form \cite{Dobaczewski2002},
 \begin{equation}
   V_{\delta}^{n/p}(\vec{r_{1}},\vec{r_{2}}) = V_{0}^{n/p}[1-\frac{1}{2}(\rho (r_{1}+ r_{2})/\rho_{0})^{\alpha}]\delta (r_{1}-r_{1})
 \end{equation}
 where the saturation density \cite{Terasaki1997} $\rho_{0}=0.16 fm^{-3}$.  Details can be obtained from the corresponding references.
\subsection{RHB Theory}
We have employed Covariant Relativistic self-consistent mean field models that are similar to Kohn–Sham
density functional theory. We constructed the Nuclear Density Functionals from Lagrangian
densities based on meson exchange and point coupling models. The pairing correlations
of nucleons are considered by the relativistic Hartree–Bogoliubov functional based on
quasi-particle operators of Bogoliubov transformations. These models are discussed briefly in this subsection.\\
\subsubsection{DD-ME Model}
The total Lagrangian \cite{Lalazissis2005} density of mesons exchange approximation which involves the isoscalar vector $\omega$ meson, the isoscalar scalar $\sigma$ meson 
and the isovector vector $\rho$ meson can be written in the following form,
\begin{eqnarray}
\label{eq:eq1}
{\cal L}  &=& \sum_i\overline{\psi_i}(i\gamma_\mu\partial^\mu-m){\psi_i}
            +\frac{1}{2} \partial_\mu\sigma\partial^\mu\sigma - \frac{1}{2}m_{\sigma}^2 \sigma^2\nonumber\\
          &&-\frac{1}{2} \Omega_{\mu\nu}\Omega^{\mu\nu}+\frac{1}{2}m_{\omega}^2\omega_{\mu}\omega^{\mu}
            -\frac{1}{4}\vec{R}_{\mu\nu} \vec{R}^{\mu\nu}+\frac{1}{2}m_{\rho}^2 \vec{\rho}_{\mu}. \vec \rho^{\,\mu}\nonumber\\
          &&-\frac{1}{4}F_{\mu\nu}F^{\mu\nu}-g_{\sigma}\overline{\psi}\psi\sigma - g_{\omega}\overline{\psi}\gamma^{\mu}\psi\omega_{\mu} 
            -g_{\rho}\overline{\psi}\vec{\tau}\gamma^{\mu}\psi.\vec{\rho}_{\mu}\nonumber\\
          &&-e\overline{\psi}\gamma^{\mu}\psi A_{\mu}.
\end{eqnarray}     

Total nuclear energy density functional of RMF for lagrangian density defined by Eq.(\ref{eq:eq1})
can be extracted in r-space as function of the Dirac spinors
$\psi,\overline{\psi}$ and the meson fields $\sigma,\omega^{\mu},\vec{\rho}^{\,\mu},A^{\mu}$;
\begin{equation}
\label{eq:eq2}
 E_{RMF}[\psi,\overline{\psi},\sigma,\omega^{\mu},\vec{\rho}^{\,\mu},A^{\mu}] = \int{d^{3}r\cal H}(r).
\end{equation}
Here, $\cal{H}$(r) Hamiltonian density can be obtained from Lagrangian density as shown by Eq.(\ref{eq:eq1}).
\subsubsection{DD-PC Model}
In a complete analogous way to meson-exchange RMF phenomenology described before, a density dependent interaction Lagrangian density 
of point coupling models \cite{Nikvsic2008}, which includes the  isoscalar-vector 
${(\overline{\psi}\gamma_{\mu}\psi)(\overline{\psi}\gamma^{\mu}\psi)}$, isoscalar-scalar ${(\overline{\psi}\psi)}^{2}$, 
and isovector-vector ${(\overline{\psi}\vec{\tau}\gamma_{\mu}\psi).{(\overline{\psi}\vec{\tau}\gamma^{\mu}\psi)}}$
four-fermion contact interactions in the isospace-space can be written as 
\begin{eqnarray}
\label{eq:eq3}
 {\cal L}&=&\overline{\psi}(i\gamma.\partial-m){\psi}
 -\frac{1}{2}\alpha_S({\rho})(\overline\psi\psi)(\overline\psi\psi)\nonumber\\
 &&-\frac{1}{2}\alpha_V(\rho)(\overline\psi\gamma^\mu\psi)(\overline\psi\gamma_\mu\psi)
-\frac{1}{2}\alpha_{TV}(\rho)(\overline\psi\vec{\tau}\gamma^\mu\psi)(\overline\psi\vec{\tau}\gamma_\mu\psi)\nonumber\\
 &&-\frac{1}{2}\delta_S(\partial_\nu\overline{\psi}\psi)(\partial^{\nu}\overline{\psi}\psi)
 -e\overline{\psi}\gamma.A\frac{1-\tau_3}{2}\psi.
 \end{eqnarray}
 The total nuclear energy density functional of RMF for the Point-Coupling Model can written as,
\begin{equation}
\label{eq:eq4}
 E_{RMF}[{\psi},\overline{\psi},A_{\mu}] = \int d^{3}r{\cal H}(r),
\end{equation}
 where $\cal{H}$(r) Hamiltonian density can be obtained from Lagrangian Density defined by Eq.(\ref{eq:eq3}).
 \subsubsection{RHB Approximation with a separable pairing interactions}
 The relativistic Hartree-Bogoliubov model \cite{Vretenar2005,Meng2006} provides an unified description of particle-particle (pp) and particle-hole
 (ph) correlations on SCMF level by using the averages of two potentials. 
 The quasiparticle operators \cite{Ring1997} are explained by unitary Bogoliubov transformation of the single-nucleon creation and annihilation operators as,
\begin{equation}
\label{eq:eq5}
 \alpha^\dagger_a = \sum_n U_{nl} c^\dagger_n + V_{nl}c_n,
\end{equation}
The relativistic Hartree-Bogoliubov energy density functional can be written as,
\begin{equation}
\label{eq:eq6}
 E_{RHB}[\rho,k] = E_{RMF}[\rho] + E_{pair}[k],
\end{equation} 
The pairing part of the relativistic Hartree-Bogoliubov functional is given as,
\begin{equation}
\label{eq:eq9}
 E_{pair}[k] = \frac{1}{4} \sum_{n_1n'_1}\sum_{n_2n'_2}k^*_{n_1n'_1} \left<{n_1n'_1\big|V^{PP}\big|n_2n'_2}\right>k_{n_2n'_2},
\end{equation}

The two body interaction matrix elements of the pairing field $\Delta$ have been computed by using a separable form of Gogny force 
introduced for hybrid RHB calculations \cite{Gonzalez1996} for spherical and deformed nuclei \cite{Nikvsic2014}. 
 The pairing energy in the nuclear ground state is given as \cite{Nikvsic2014},
\begin{equation}
\label{eq:eq15}
 E_{pair} = -G\sum_{N} P^{*}_{N}P_{N}.
\end{equation}
The details can be found in the corresponding references. 
\section{Results And Discussions}
We present our results for Binding Energies,
shell closure parameter $D_{n} (N)$ and the differential variation of the two-neutron separation energy $dS_{2n} (Z, N )$ for the isotopic chains of Si, S, Ar and Ca nuclides. The theoretical calculations  are carried out within the framework of HFB theory and RHB theory discussed briefly in the previous section.  HFB Calculations are carried out for different skyrme parameterizations by using the HFB code \cite{Perez2017} with harmonic oscillator basis. RHB calculations are computed with DIRHB code \cite{Nikvsic2014}. 
\subsection{Binding Energies}
\begin{figure}
\centering
\includegraphics[trim=0 0 0 0,clip,scale=0.4]{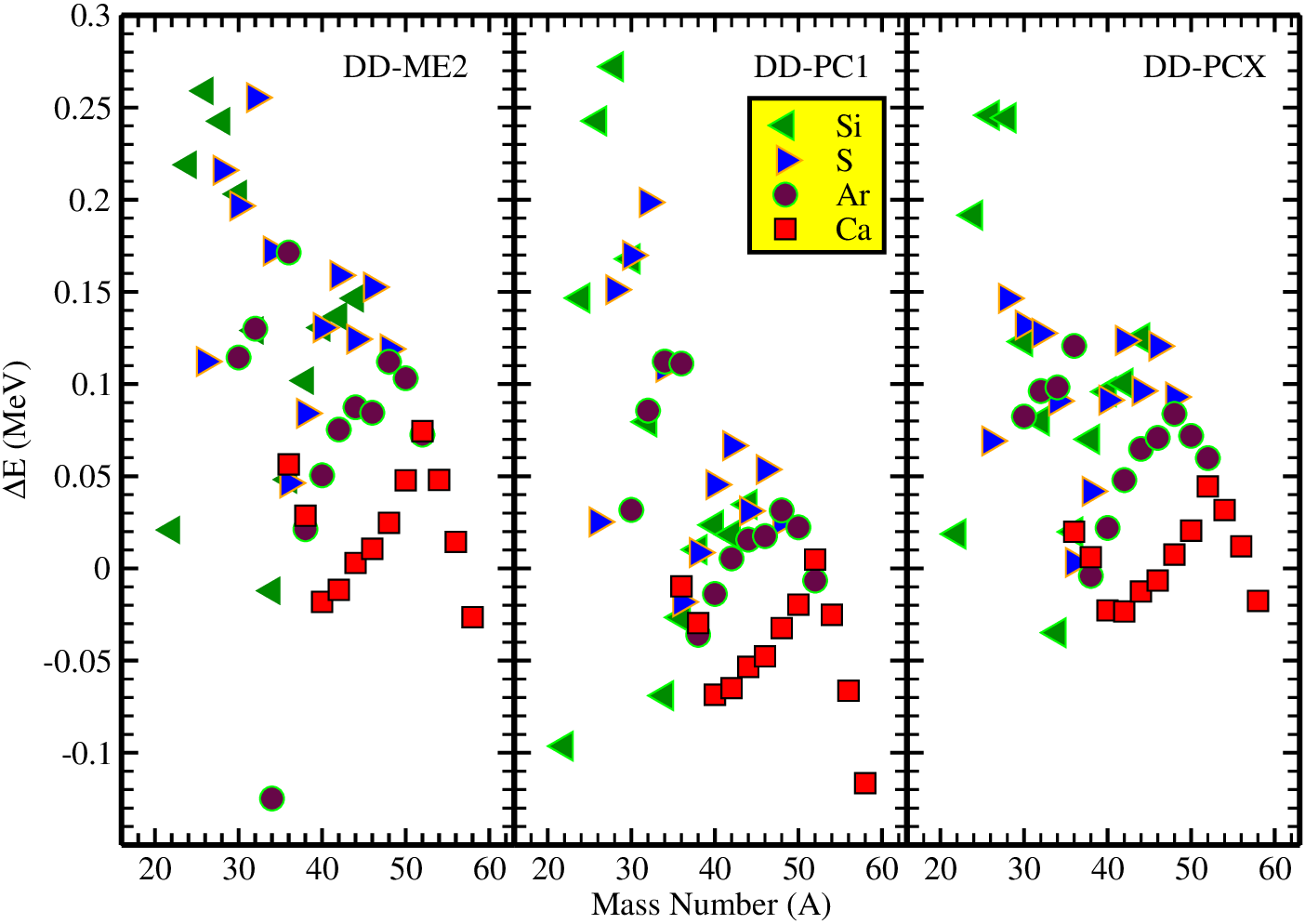}
\caption{\label{Fig1}(color online) $\Delta E$ in units of MeV representing the difference in experimental \cite{Wang2017} and theoretical binding energy per nucleon, plotted as a function of mass number A for the exotic nuclei of Silicon, Sulphur, Argon and Calcium. The theoretical estimates are computed by using Relativistic-Hartree-Bogoliubov  density functional based on Meson Exchange model parameters DD-ME2 (leftmost panel), Point Coupling model parameters DD-PC1 (middle panel) 
and Point Coupling model parameters DD-PCX (right most panel).}
\end{figure}

\begin{figure}
\centering
\includegraphics[trim=0 0 0 0,clip,scale=0.4]{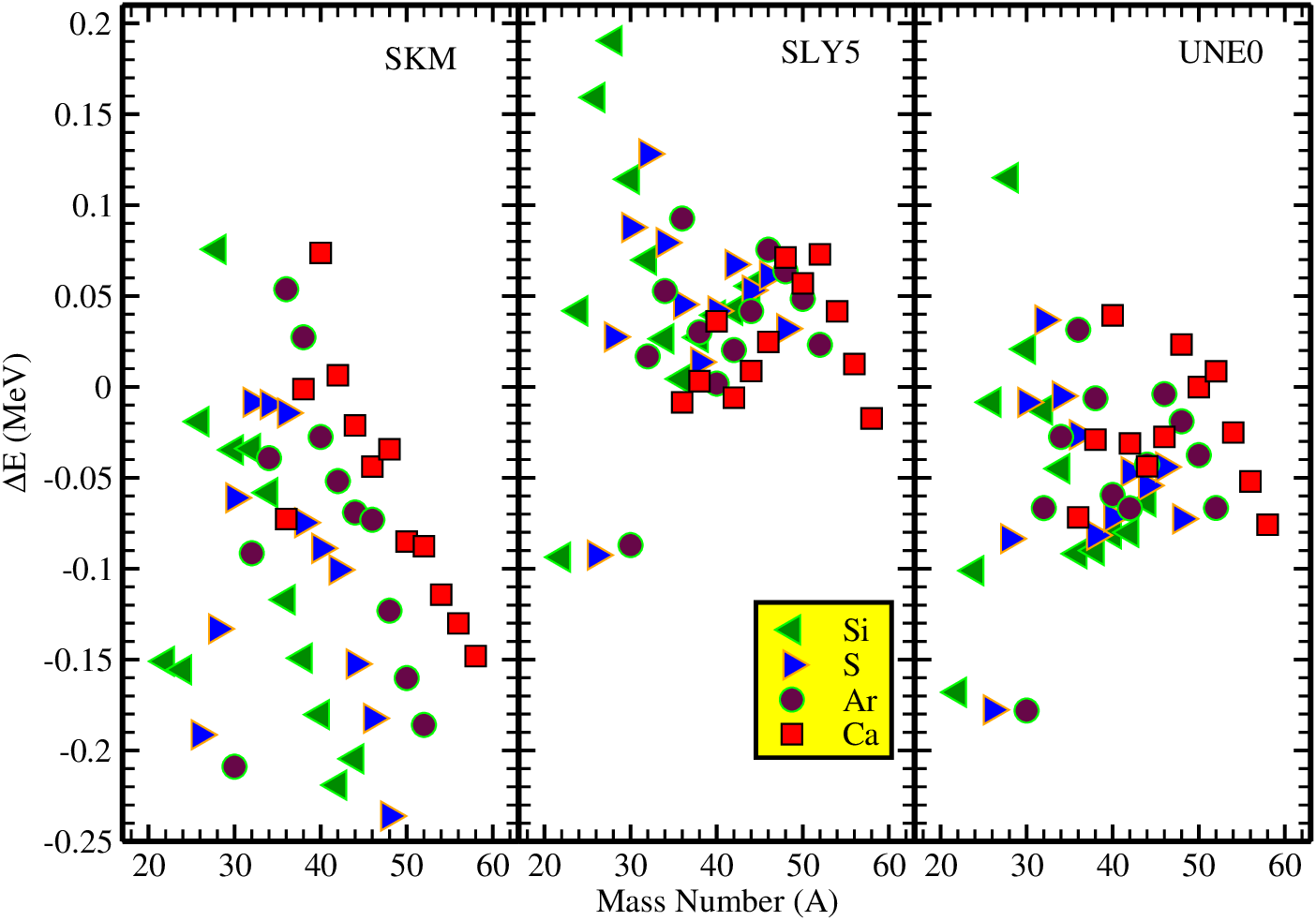}
\caption{\label{Fig2}(color online) $\Delta E$ in units of MeV representing the difference in experimental \cite{Wang2017} and theoretical binding energy per nucleon, plotted as a function of mass number A for the exotic nuclei of Silicon, Sulphur, Argon and Calcium. The theoretical estimates are computed by using Hartree-Fock-Bogoliubov  density functional based on Skyrme forces parameters SKM* (leftmost panel),     SLY5 (middle panel) 
and UNEDF0 (right most panel).}
\end{figure}

Binding Energy is the underlying property of the nuclides which provides the deep insights about the nuclear structure. Binding Energy refers to that energy which when given to the nuclei, break it apart into its constituent nucleons. We have computed the theoretical results of  Binding Energy per nucleon (BE/A) for  the isotopic chains of the light to middle weight nuclides of Si, S, Ar and Ca. Theoretical  Computation is done with the Relativistic-Hartree-Bogoliubov and Hartree-Fock-Bogoliubov Nuclear Density Functional Theory. Figure (\ref{Fig1} and \ref{Fig2}) represents the variation of the quantity $\Delta E$ against the mass number A of particular nuclides. The quantity $\Delta E$ is defined as the difference between the Experimental Binding Energy \cite{Wang2017} per nucleon and the Theoretical Binding Energy per nucleon.
\begin{equation}
 \label{eq:eq101}
 \Delta E = \left(\frac{BE}{A}\right)_{Exp}- \left(\frac{BE}{A}\right)_{The}
 \end{equation}
 The theoretical results of binding
energy shown in Figure (\ref{Fig1}) are calculated by using the relativistic nuclear density functional based on
meson exchange model parameters DD-ME2 \cite{Lalazissis2005} (leftmost panel) and point coupling model parameters  DD-PC1 \cite{Yuksel2019} (middle panel)  and DD-PC1 \cite{Nikvsic2008} (right most panel).
Figure (\ref{Fig2}) represents our theoretical estimates of BE/A extracted with the Hartree-Fock-Bogoliubov Theory with Skyrme parameterizations UNEDF0 \cite{Kortelainen2010} \cite{Chabanat1997} (lefmost panel), SLY5 (middle panel) and SKM* \cite{Dobaczewski1984} (rightmost panel). Our theoretical results are matching reasonably with the experimental results available \cite{Wang2017}. Negative values of the quantity $\Delta E$ defined in equation (\ref{eq:eq101}) shows that the theoretical estimates just exceeds the experimental data extracted from ref. \cite{Wang2017}. With the Relativistic-Hartree-Bogoliubov theory shown in Figure (\ref{Fig1}), the absolute value quantity $\Delta E$ varies from 
0.01 to 0.26 for Si isotopes, 0.01 to 0.25 for S isotopes, 0.01 to 0.17 for Ar isotopes and 0.01 to 0.11 for the Ca isotopes. Absolute value of  $\Delta E$ varies from 0.01 to 0.22 for Si isotopes, 0.01 to 0.23 for S isotopes, 0.01 to 0.21 for Ar isotopes and 0.01 to 0.15 for the Ca isotopes when the results are calculated with Hartree-Fock-Bogoliubov theory based on skyrme parameterizations as shown in Figure (\ref{Fig2}).
\subsection{Single-neutron separation energy and Shell closure parameter} 
\begin{figure}
\centering
\includegraphics[trim=0 0 0 0,clip,scale=0.4]{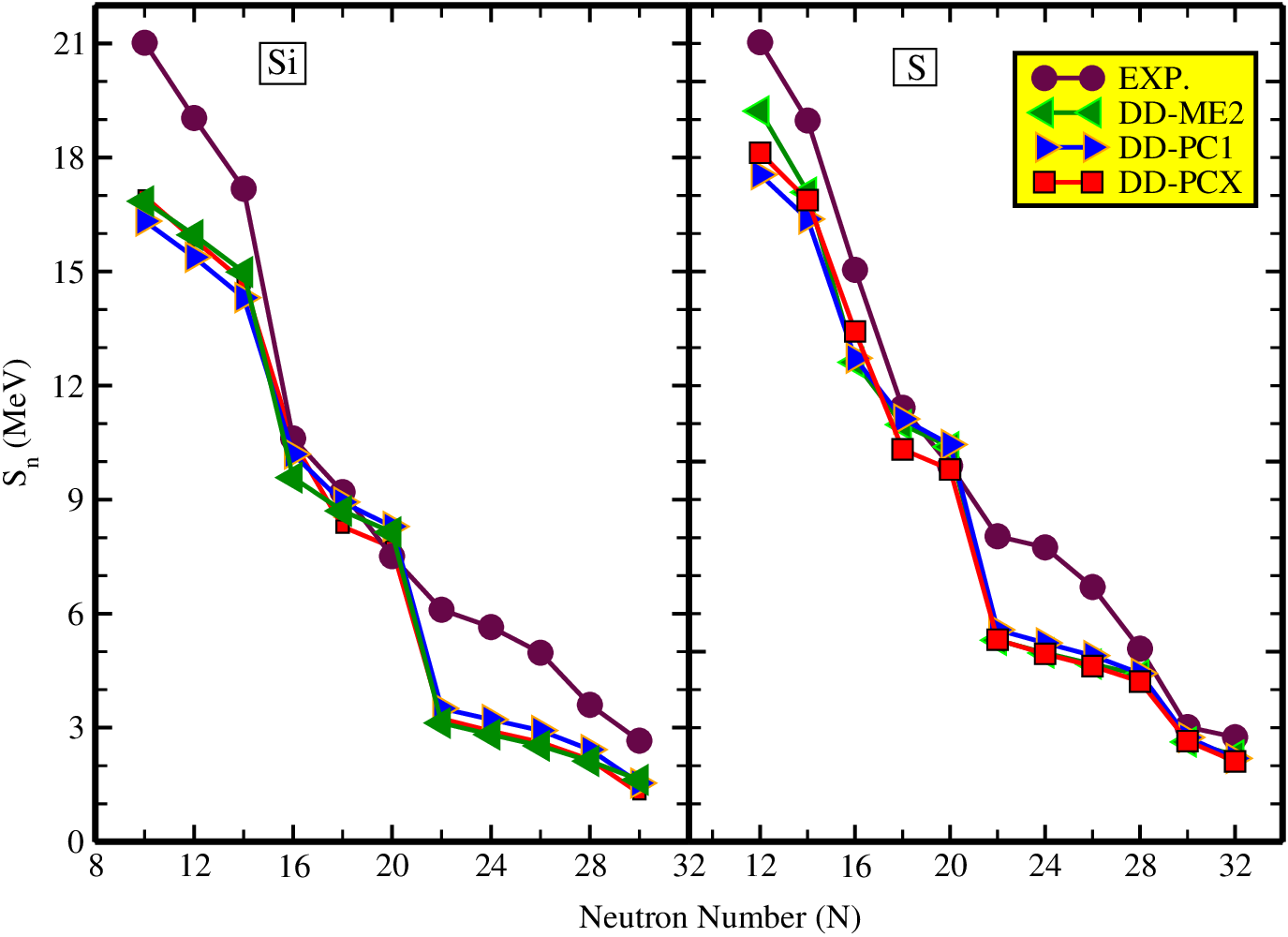}
\caption{\label{Fig3} (color online) The variation in experimental \cite{Wang2017} and theoretical single-neutron separation energy $S_{n}(Z,N)$ in units of MeV, 
plotted as a function of neutron number N, for the exotic nuclei of Silicon (left panel) and Sulphur (right panel). 
The theoretical estimates are computed by using Relativistic-Hartree-Bogoliubov  density functional based on Meson Exchange model parameters DD-ME2, Point Coupling model parameters DD-PC1 
and Point Coupling model parameters DD-PCX.}
\end{figure}

\begin{figure}
\centering
\includegraphics[trim=0 0 0 0,clip,scale=0.4]{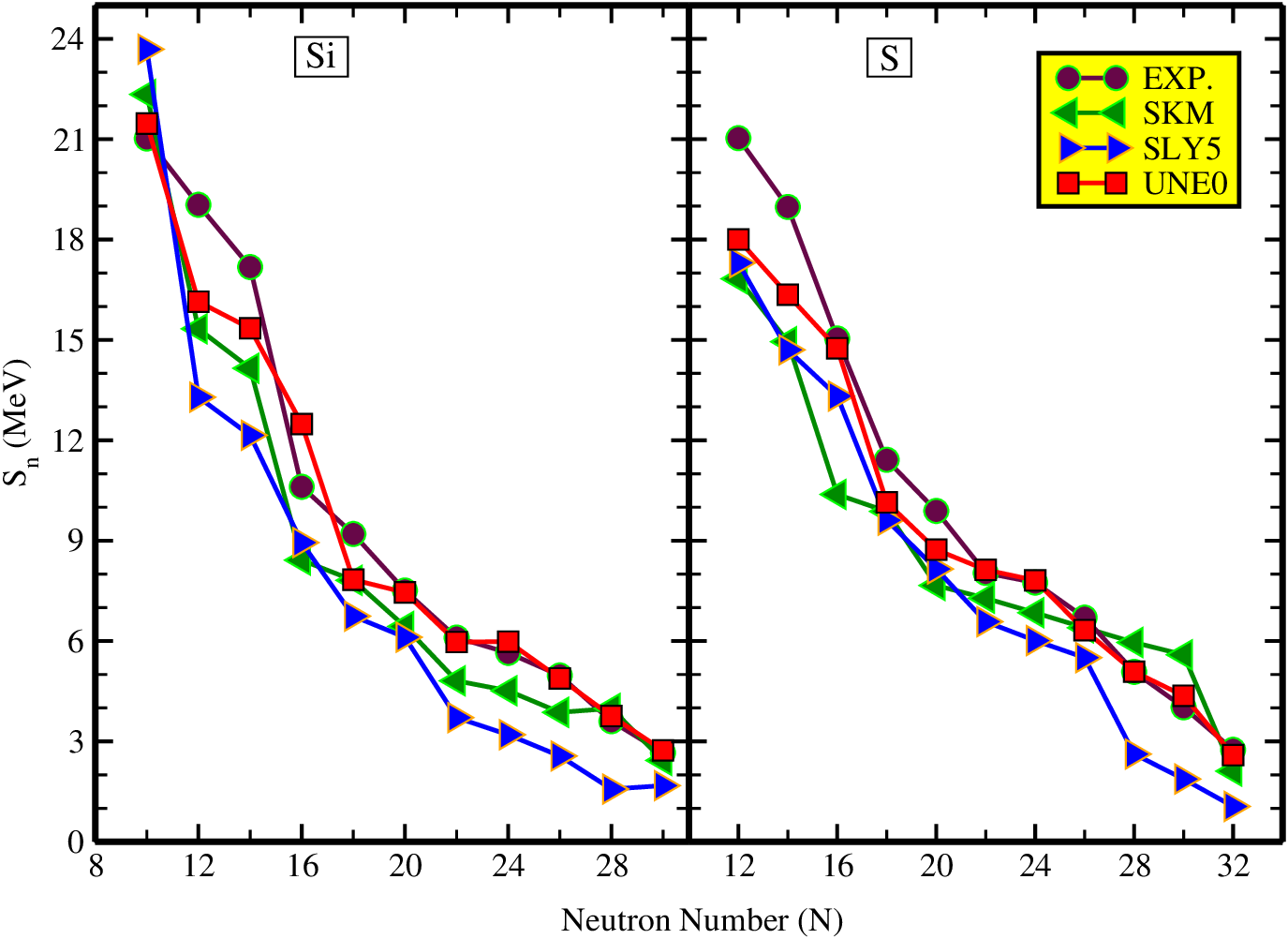}
\caption{\label{Fig4} (color online) The variation in experimental \cite{Wang2017} and theoretical single-neutron separation energy $S_{n}(Z,N)$ in units of MeV, 
plotted as a function of neutron number N, for the exotic nuclei of Silicon (left panel) and Sulphur (right panel). 
The theoretical estimates are computed by using Hartree-Fock-Bogoliubov  density functional based on Skyrme forces parameters SKM*,     SLY5 and UNEDF0.}
\end{figure}

\begin{figure}
\centering
\includegraphics[trim=0 0 0 0,clip,scale=0.4]{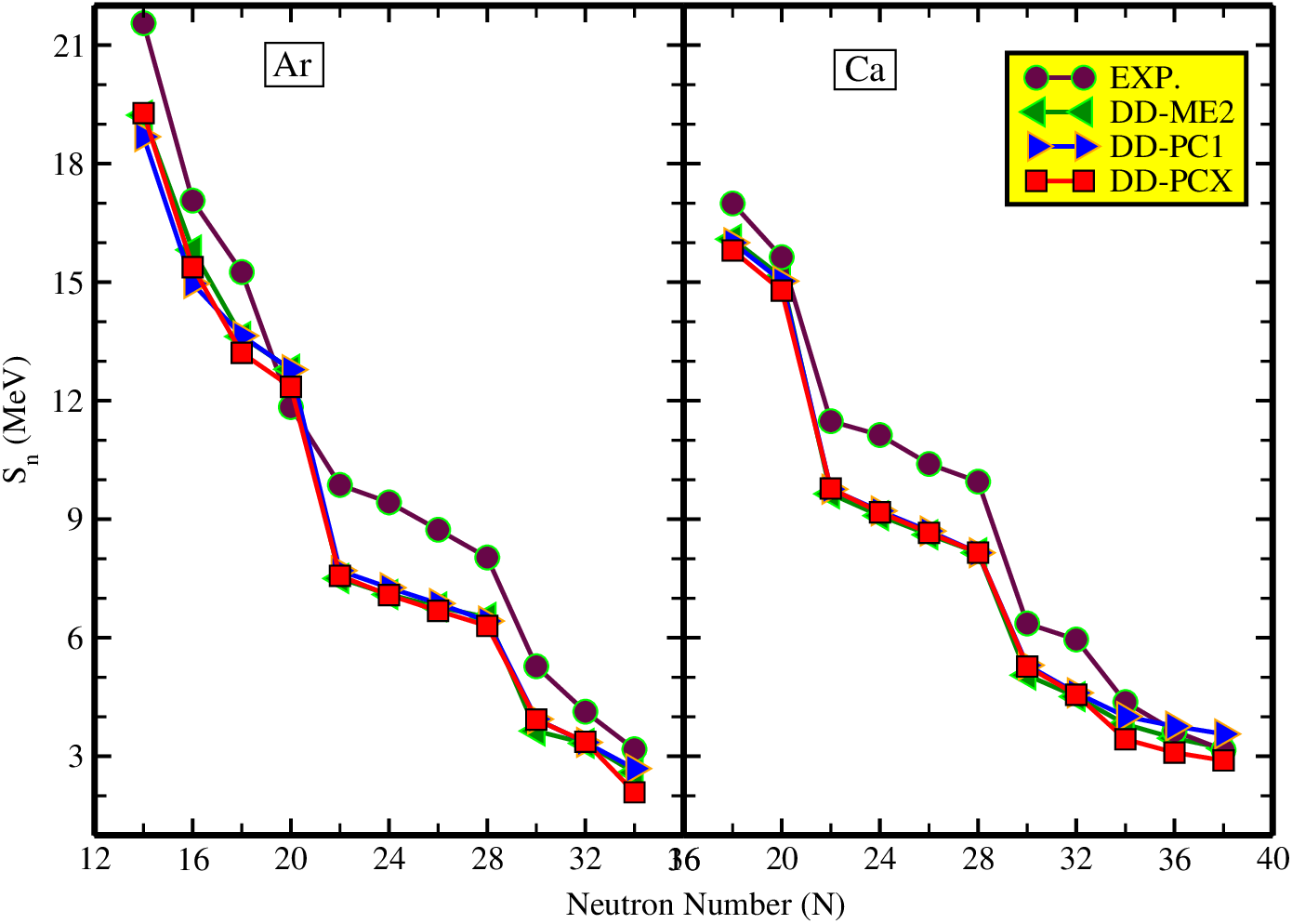}
\caption{\label{Fig5} (color online) The variation in experimental \cite{Wang2017} and theoretical single-neutron separation energy $S_{n}(Z,N)$ in units of MeV, 
plotted as a function of neutron number N, for the exotic nuclei of Argon (left panel) and Calcium (right panel). 
The theoretical estimates are computed by using Relativistic-Hartree-Bogoliubov  density functional based on Meson Exchange model parameters DD-ME2, Point Coupling model parameters DD-PC1 
and Point Coupling model parameters DD-PCX.}
\end{figure}

\begin{figure}
\centering
\includegraphics[trim=0 0 0 0,clip,scale=0.4]{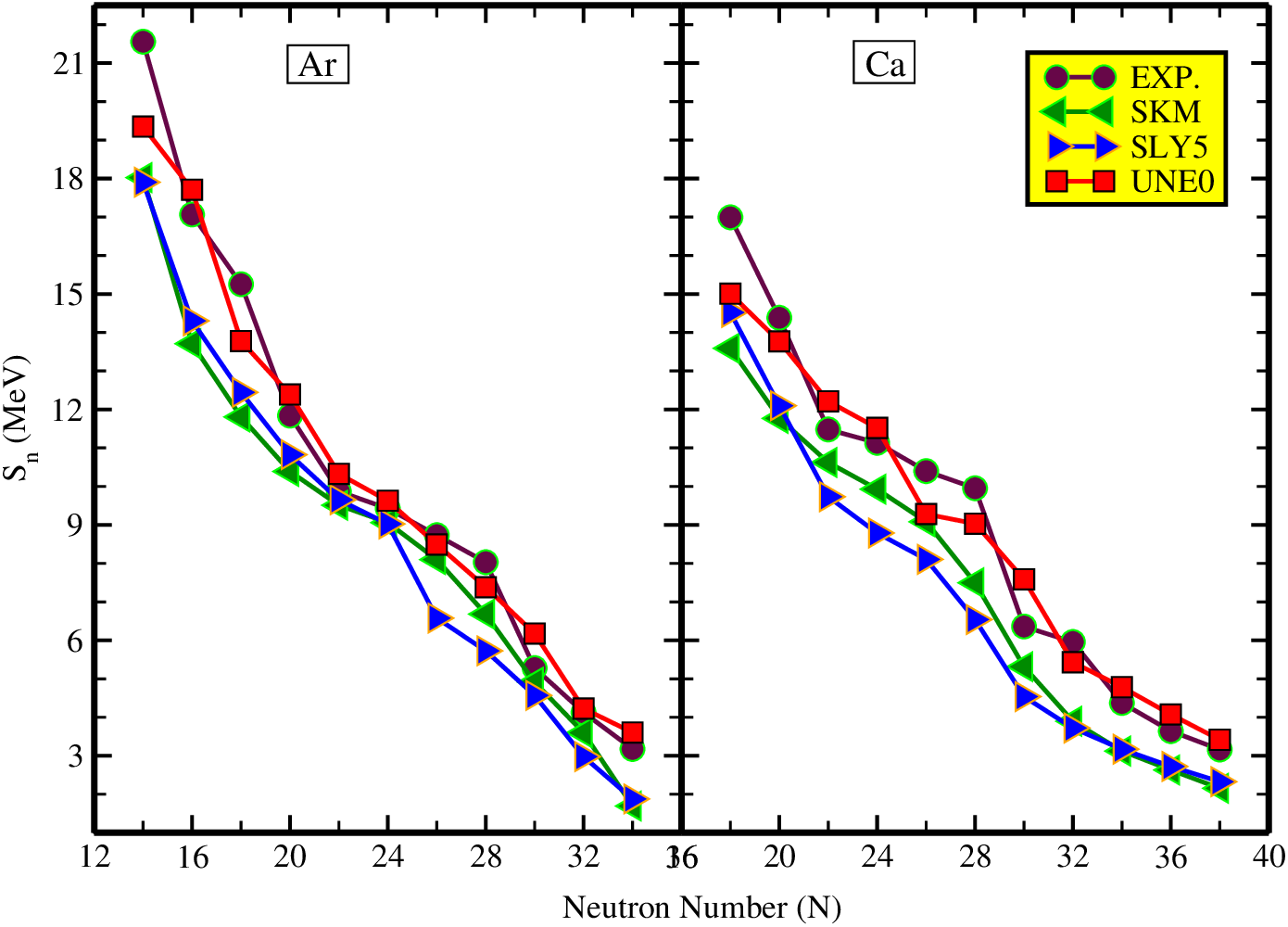}
\caption{\label{Fig6} (color online) The variation in experimental \cite{Wang2017} and theoretical single-neutron separation energy $S_{n}(Z,N)$ in units of MeV, 
plotted as a function of neutron number N, for the exotic nuclei of Argon (left panel) and Calcium (right panel). 
The theoretical estimates are computed by using Hartree-Fock-Bogoliubov  density functional based on Skyrme forces parameters SKM*,     SLY5 and UNEDF0.}
\end{figure}

\begin{figure}
\centering
\includegraphics[trim=0 0 0 0,clip,scale=0.4]{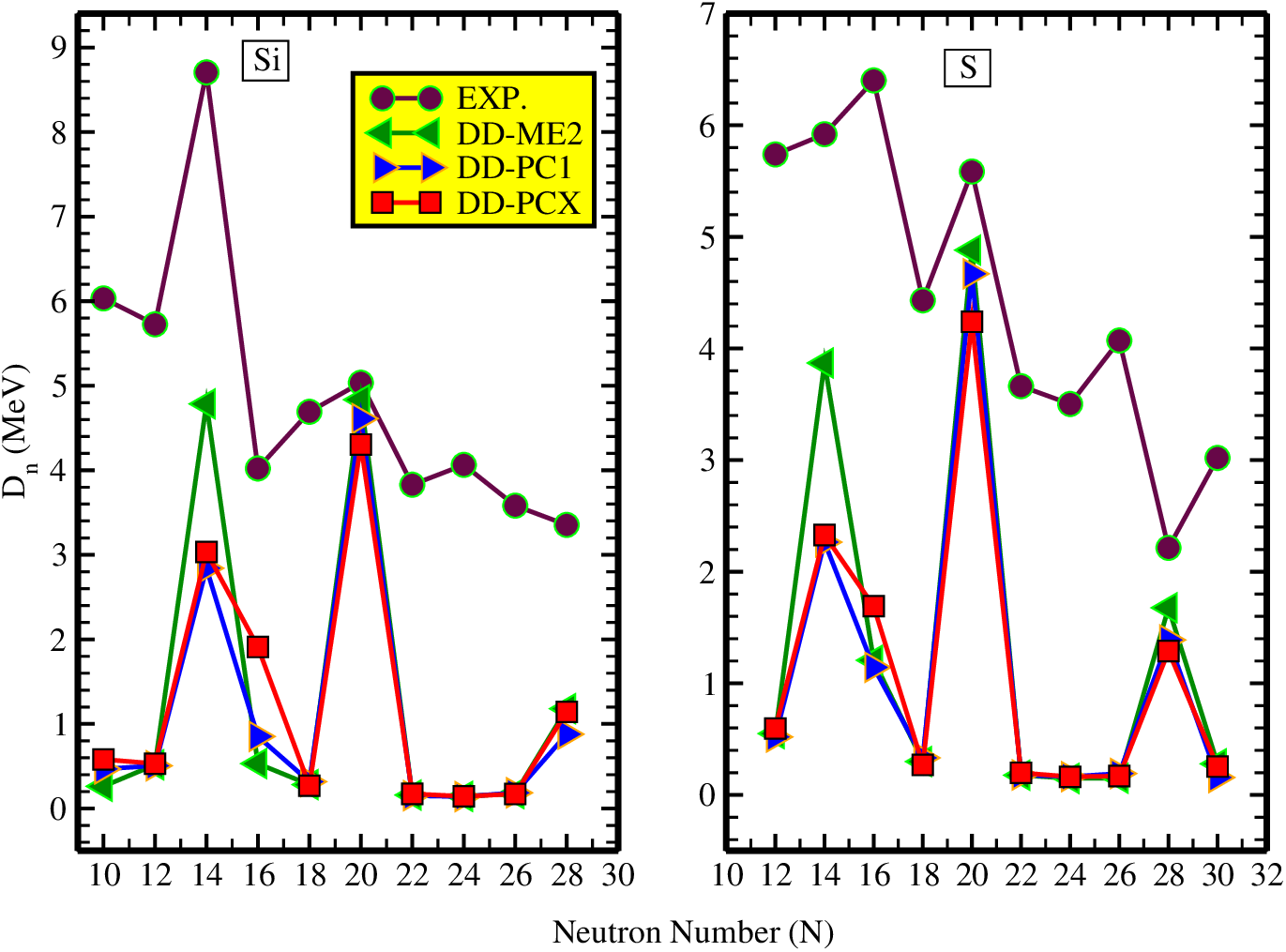}
\caption{\label{Fig7}(color online)  $D_{n}(Z,A)$ in units of MeV, plotted as a function of neutron number N, 
for the exotic nuclei of Silicon (left panel) and Sulphur (right panel). The theoretical estimates are computed by using Relativistic-Hartree-Bogoliubov  density functional based on Meson Exchange model parameters DD-ME2, Point Coupling model parameters DD-PC1 
and Point Coupling model parameters DD-PCX. Experimental data is taken from \cite{Wang2017}.}
\end{figure}

\begin{figure}
\centering
\includegraphics[trim=0 0 0 0,clip,scale=0.4]{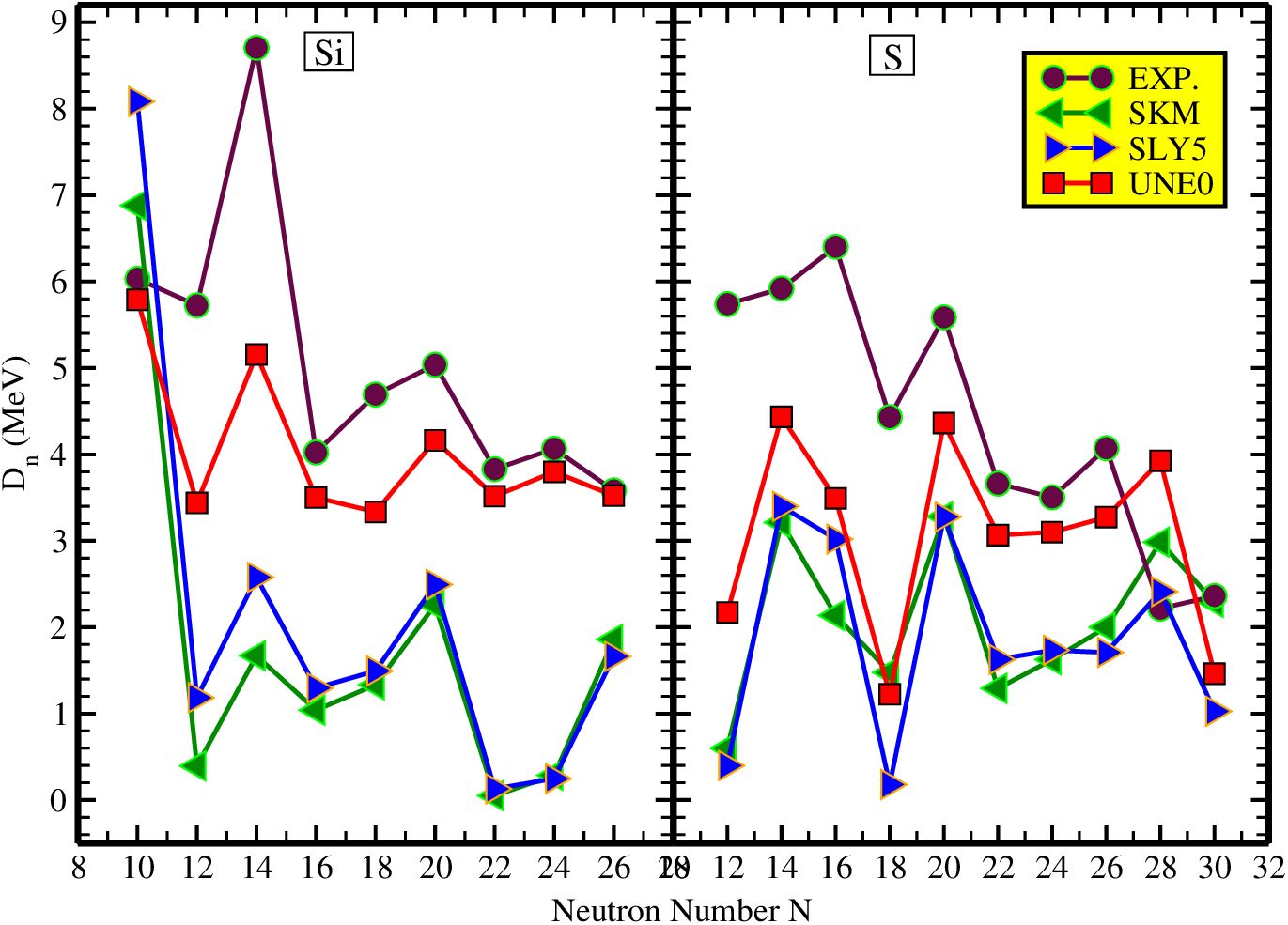}
\caption{\label{Fig8}(color online)  $D_{n}(Z,A)$ in units of MeV, plotted as a function of neutron number N, 
for the exotic nuclei of Silicon (left panel) and Sulphur (right panel). The theoretical estimates are computed by using Hartree-Fock-Bogoliubov  density functional based on Skyrme forces parameters SKM*,     SLY5 and UNEDF0. Experimental data is taken from \cite{Wang2017}.}
\end{figure}

\begin{figure}
\centering
\includegraphics[trim=0 0 0 0,clip,scale=0.4]{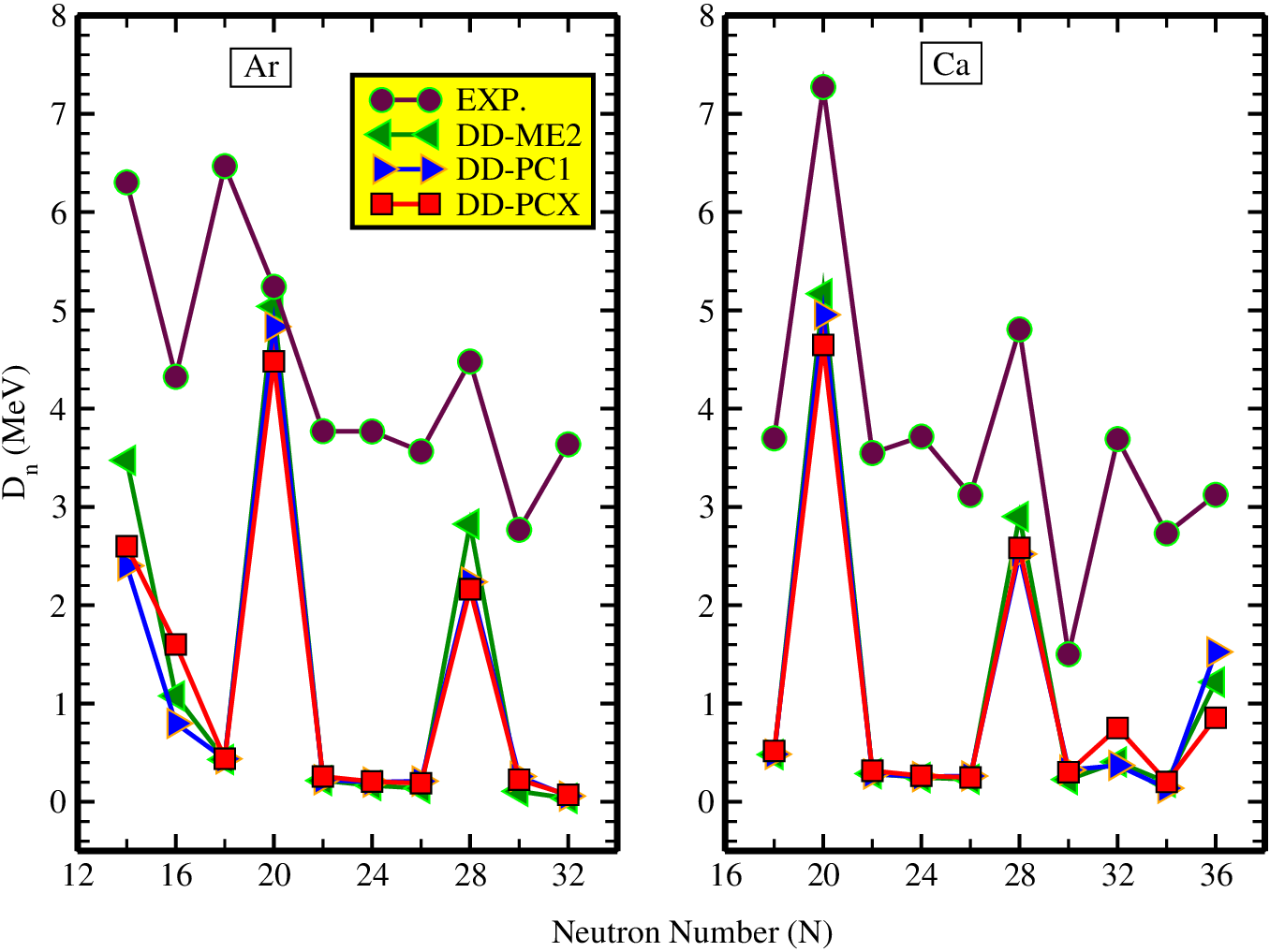}
\caption{\label{Fig9}(color online)  $D_{n}(Z,A)$ in units of MeV, plotted as a function of neutron number N, 
for the exotic nuclei of Argon (left panel) and Calcium (right panel). The theoretical estimates are computed by using Relativistic-Hartree-Bogoliubov  density functional based on Meson Exchange model parameters DD-ME2, Point Coupling model parameters DD-PC1 
and Point Coupling model parameters DD-PCX. Experimental data is taken from \cite{Wang2017}.}
\end{figure}

\begin{figure}
\centering
\includegraphics[trim=0 0 0 0,clip,scale=0.4]{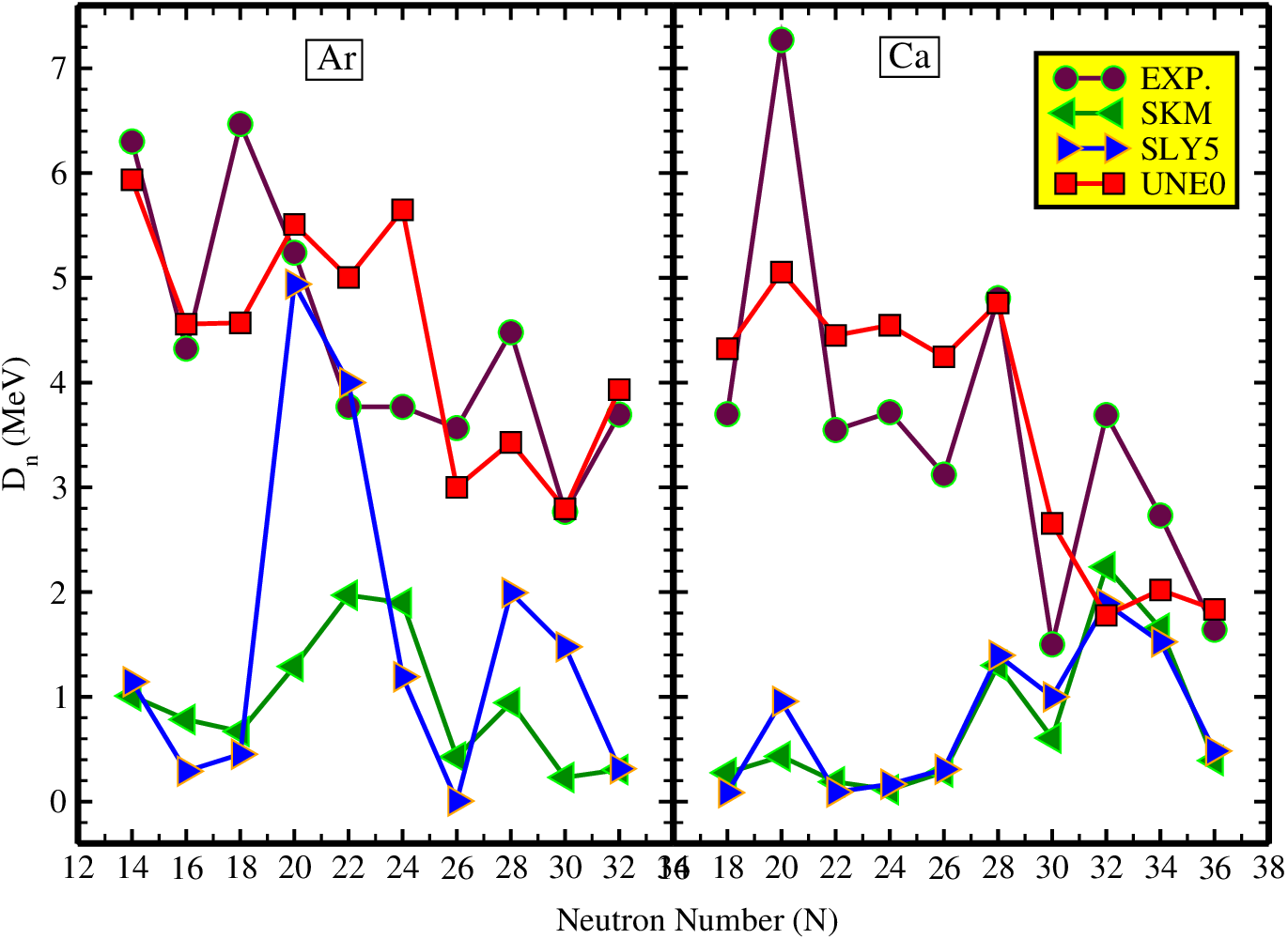}
\caption{\label{Fig10}(color online)  $D_{n}(Z,A)$ in units of MeV, plotted as a function of neutron number N, 
for the exotic nuclei of Argon (left panel) and Calcium (right panel). The theoretical estimates are computed by using Hartree-Fock-Bogoliubov  density functional based on Skyrme forces parameters SKM*,     SLY5 and UNEDF0. Experimental data is taken from \cite{Wang2017}.}
\end{figure}

Single-neutron separation energy denoted as $S_{n}(Z,N)$ (Z=proton number and N=neutron number) is an important property of nuclei which acts as probe to identify the signature of shell closures or its collapse by investigating the recently proposed physical quantity $D_{n}$(N) \cite{Brown2013}.
The quantity $S_{n}(Z,N)$ is defined as
\begin{equation}
\label{eq:Sn}
 S_n(Z,A) = B(Z,N)-B(Z,N-1)
\end{equation}
where $B(Z,N)$ represents the binding energy of the nuclei with atomic number Z and neutron number N. The physical quantity D$_{n}$(Z,A) representing the shell closure
is calculated using the formula \cite{Brown2013},
\begin{equation}
\label{eq:Dn}
D_n(Z,N) = {(-1)}^{N+1}[S_n(Z,N+1)-S_n(Z,N)]
\end{equation}
In Figures(\ref{Fig3},\ref{Fig4},\ref{Fig5}, and \ref{Fig6}), we present our results of Single-neutron separation energy S$_{n}$(Z,N) in MeV plotted as a function of neutron number N, for the chain of exotic isotopic nuclei in Si, S, Ar and Ca. The theoretical results are extracted with RHB and HFB model based on different parameterizations mentioned in the individuals Figures and their respective captions. The experimental 
values of S$_{n}$(Z,N) are computed from the data of single neutron separation energies from \cite{Wang2017}. Our estimates of $S_n$(Z,N) are in good agreement with the experimental ones. Figures(\ref{Fig7},\ref{Fig8},\ref{Fig9}, and \ref{Fig10}), represents our results of  theoretical shell closures D$_{n}$(Z,N) in MeV plotted as a function of neutron number N, for the isotopic chain of nuclei in Si, S, Ar and Ca (see caption of figures for required details).
As reported in ref. \cite{42}, the isotopes of Si varying from $^{28}$Si
to $^{34}$Si, exhibit relatively high E$(2^+_1) = 3.33$ MeV energies in $^{34}$Si nucleus which indicates  the Z = 14 as the new subshell gap with N = 20. However, it is also interesting to see that the magnitudes 
of  E$(2^+_1)$ energies gradually decreases from 1.4 Mev to 0.74 MeV in $^{36}$Si to $^{42}$Si. This suggest and indicate the collapse of the N = 28 shell closure in $^{42}$Si nucleus \cite{Bastin2007}. 
In the left panels of Figures (\ref{Fig7} and \ref{Fig8}), we observe the 
peaks in the D$_{n}$(Z,N) at neutron number, N = 14 and 20, which gives the presence of shell closures. Afterwards, there is decrease in values of  
D$_{n}$(Z,N)  upto N = 30. In right panels of Figures (\ref{Fig7} and \ref{Fig8} ),  we present the results of D$_{n}$(Z,N) for the isotopic chain of Sulphur.
We observed  peaks in theoretical  D$_{n}$(Z,N) values at neutron numbers, N = 14 and 20 and, a small peak at N =  28 showing the weak shell closures. D$_{n}$(Z,N) val
ues Extracted experimentally \cite{Wang2017} represent peak at N = 20 only with decrease  in D$_{n}$(Z,N) values before and after 
N = 20. The experimentally extracted D$_{n}$(Z,N) values has a peak at N = 16 in $^{32}$S nucleus signifies the sub-shell gap at neutrons filled
1s$_{1/2}$ orbit in 0d-1s Shell orbitals.  Recent experimental measurements \cite{Gaudefroy2009}
of neutrons in $\nu 1p_{3/2}$ orbitals in ground state of $^{44}$S nucleus in $0f-1p$ valence space fortifies the collapse of N = 28 shell gap.
The Figures (\ref{Fig9} and \ref{Fig10}) represents the variation
in D$_{n}$(Z,A) as a function of neutron number for the isotopic chain Argon (shown in left panels) and Calcium (shown in right panels). 
The theoretical and experimentally extracted D$_{n}$(Z,A) values are exhibiting peaks at neutron numbers N =  20 and 28 in the isotopes 
of Argon and Calcium. 
These studies suggest new shell structure at Z or N = 14 due to filled 0d$_{5/2}$ orbit of $0d_{5/2}$ - 0d$_{3/2}$ spin-orbit partner 
in $^{28}$Si, $^{30}$S, $^{32}$Ar and $^{34}$Ca nuclei in ground state. Experimental data also fortifies our theoretical estimates.
\subsection{Differential Variation of Two Neutron Separation Energy}
\begin{figure}
\centering
\includegraphics[trim=0 0 0 0,clip,scale=0.4]{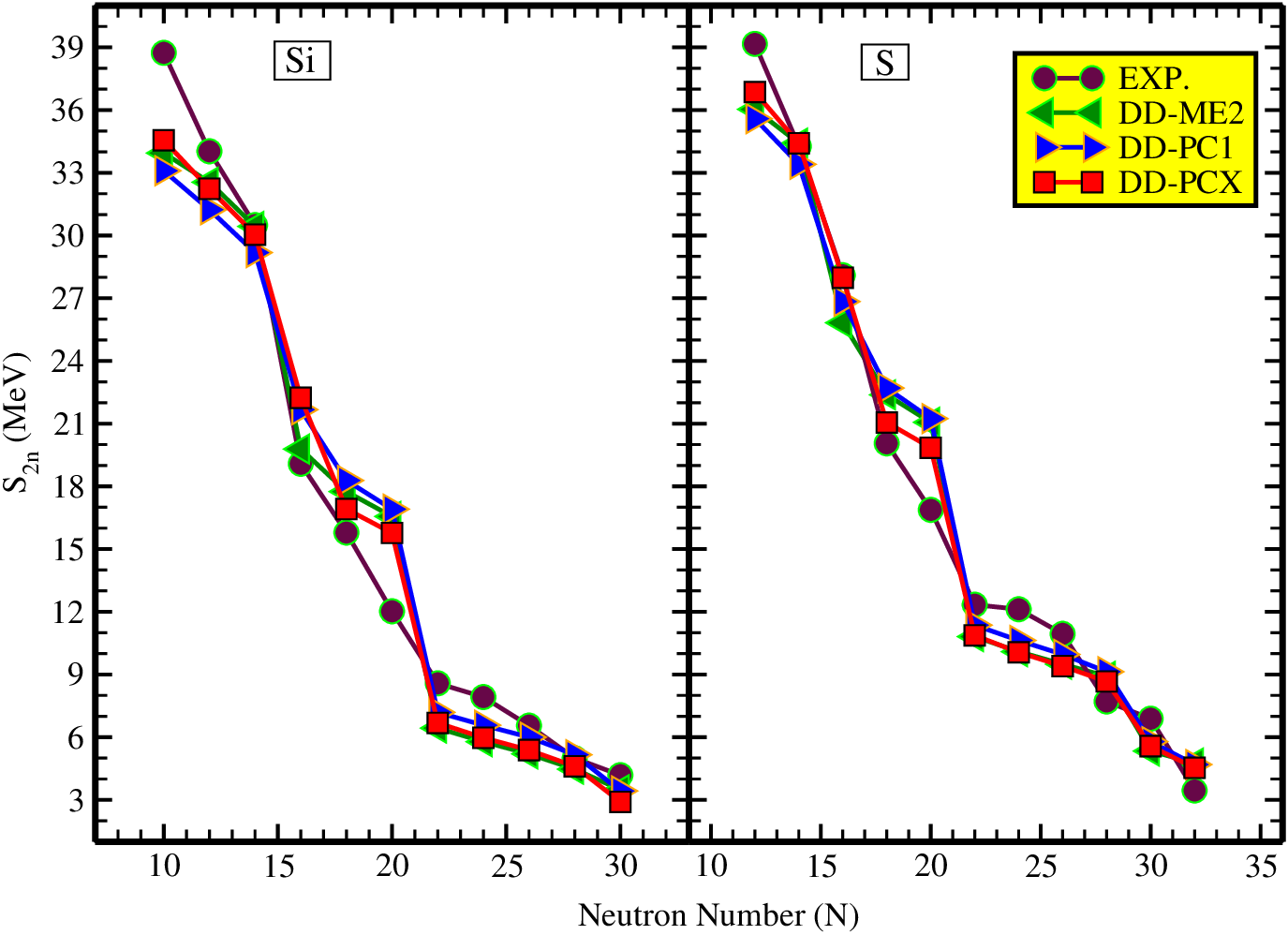}
\caption{\label{Fig11} (color online) The variation in experimental \cite{Wang2017} and theoretical two-neutron separation energy $S_{2n}(Z,N)$ in units of MeV, 
plotted as a function of neutron number N, for the exotic nuclei of Silicon (left panel) and Sulphur (right panel). 
The theoretical estimates are computed by using Relativistic-Hartree-Bogoliubov  density functional based on Meson Exchange model parameters DD-ME2, Point Coupling model parameters DD-PC1 
and Point Coupling model parameters DD-PCX.}
\end{figure}

\begin{figure}
\centering
\includegraphics[trim=0 0 0 0,clip,scale=0.4]{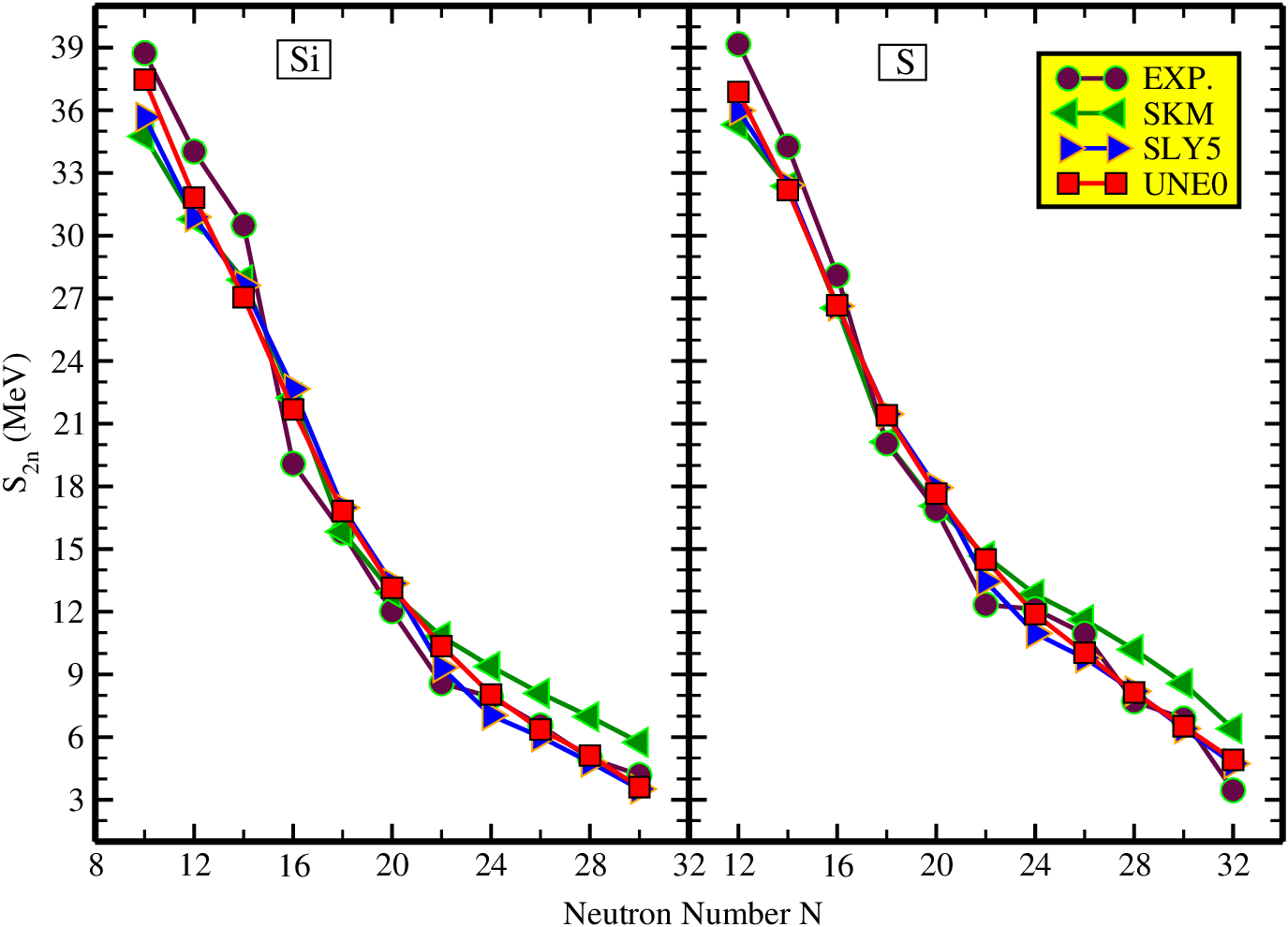}
\caption{\label{Fig12} (color online) The variation in experimental \cite{Wang2017} and theoretical single-neutron separation energy $S_{2n}(Z,N)$ in units of MeV, 
plotted as a function of neutron number N, for the exotic nuclei of Silicon (left panel) and Sulphur (right panel). 
The theoretical estimates are computed by using Hartree-Fock-Bogoliubov  density functional based on Skyrme forces parameters SKM*,     SLY5 and UNEDF0.}
\end{figure}

\begin{figure}
\centering
\includegraphics[trim=0 0 0 0,clip,scale=0.4]{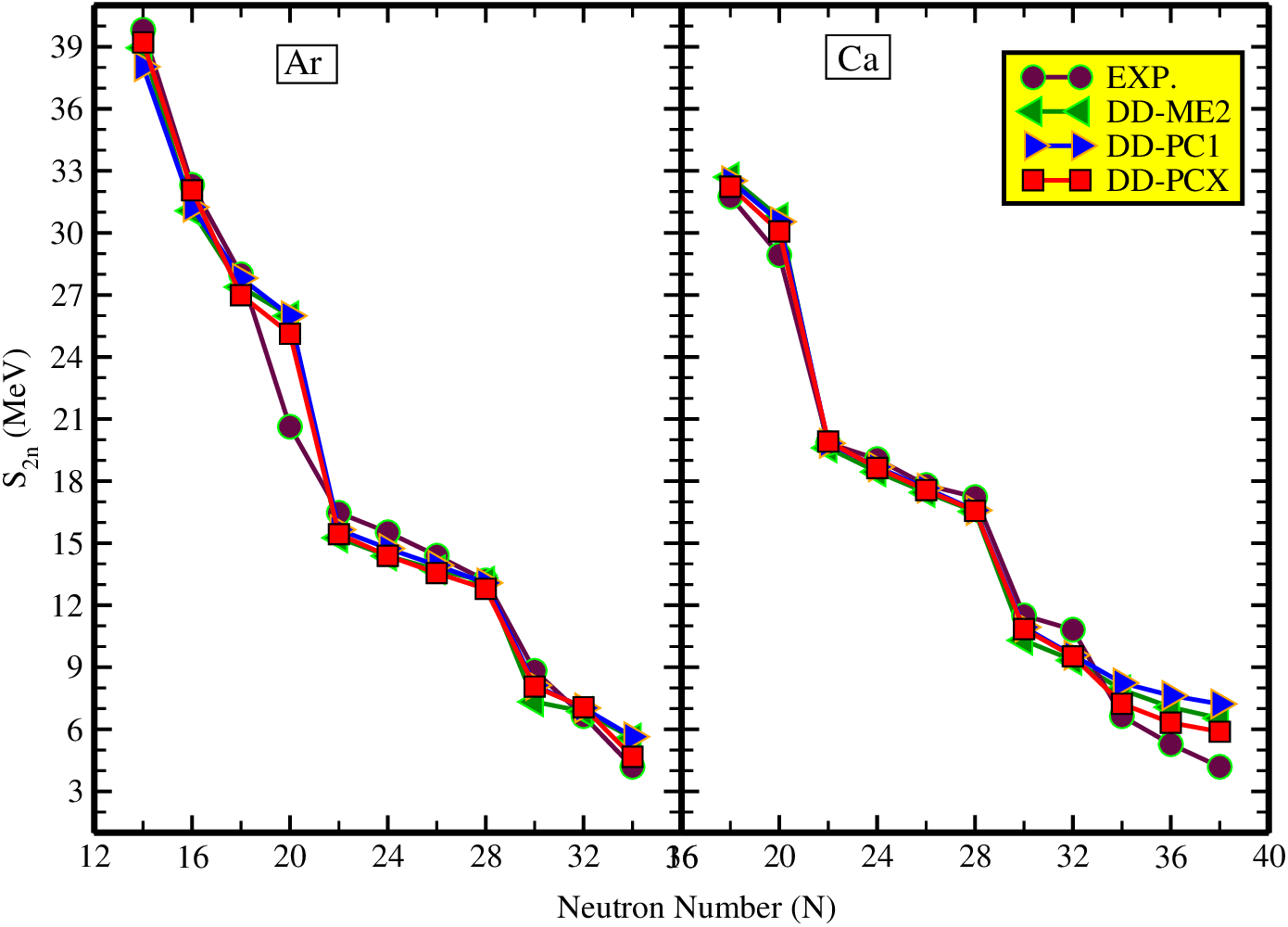}
\caption{\label{Fig13} (color online) The variation in experimental \cite{Wang2017} and theoretical single-neutron separation energy $S_{2n}(Z,N)$ in units of MeV, 
plotted as a function of neutron number N, for the exotic nuclei of Argon (left panel) and Calcium (right panel). 
The theoretical estimates are computed by using Relativistic-Hartree-Bogoliubov  density functional based on Meson Exchange model parameters DD-ME2, Point Coupling model parameters DD-PC1 
and Point Coupling model parameters DD-PCX.}
\end{figure}

\begin{figure}
\centering
\includegraphics[trim=0 0 0 0,clip,scale=0.4]{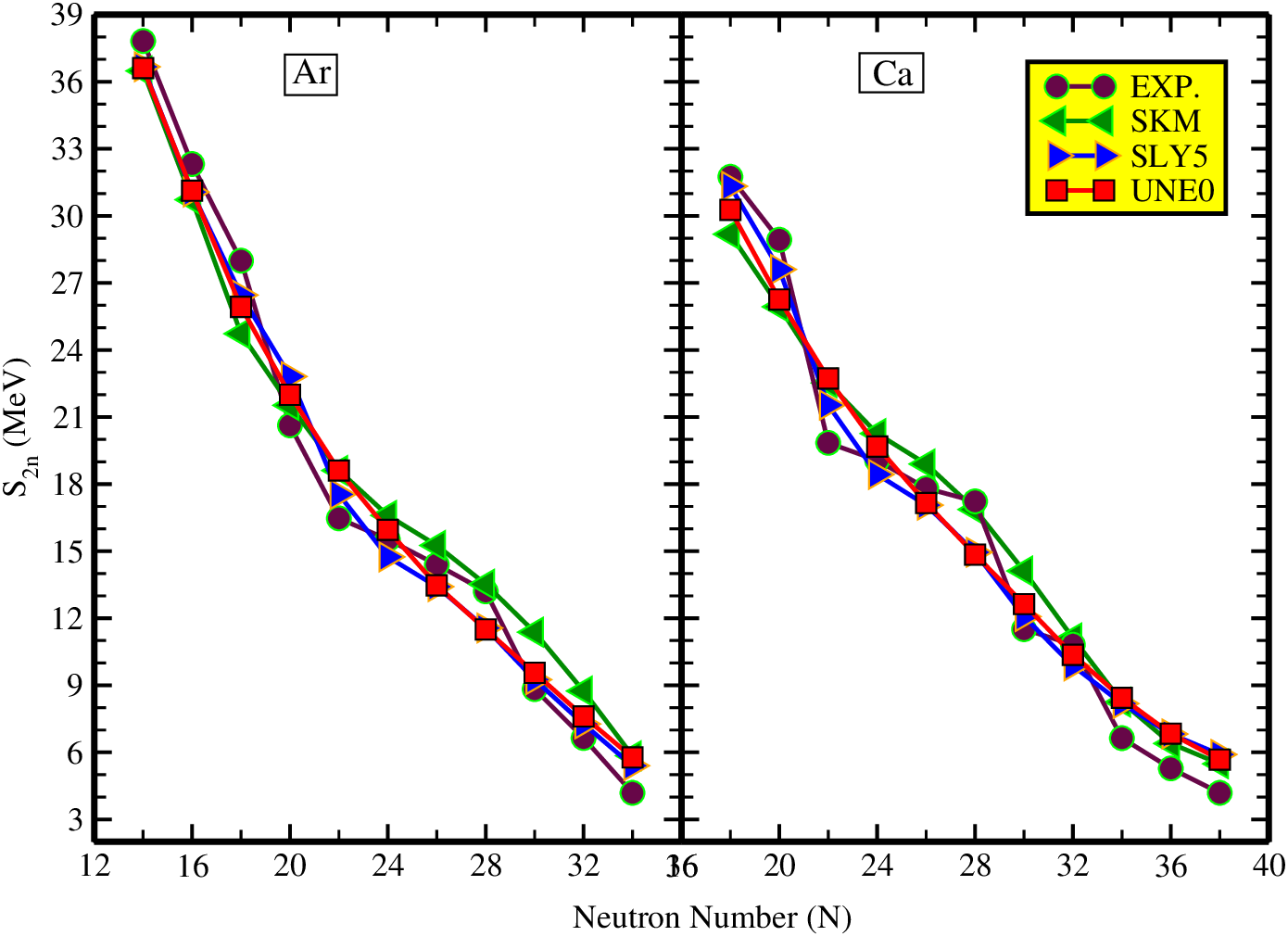}
\caption{\label{Fig14} (color online) The variation in experimental \cite{Wang2017} and theoretical single-neutron separation energy $S_{2n}(Z,N)$ in units of MeV, 
plotted as a function of neutron number N, for the exotic nuclei of Argon (left panel) and Calcium (right panel). 
The theoretical estimates are computed by using Hartree-Fock-Bogoliubov  density functional based on Skyrme forces parameters SKM*,     SLY5 and UNEDF0.}
\end{figure}

Just like the Single-neutron separatin energy $S_{n}$(Z,N) defined in Equation \ref{eq:Sn}, the another term two-neutron separation energy $S_{2n}$(Z,N) is equally important. In the given atomic mass region, the signature for the appearance and collapse of shell closures is readily identified by examining the recently proposed physical quantity $dS_{2n}$(Z,N) \cite{Bhuyan2015} based on two-neutron separation energies. The two-neutron separation energy is defined as the energy required to remove two neutrons from a nucleus. It is calculated by using the formula,
\begin{equation}
\label{eq:S2n}
 S_{2n}(Z,N) = [B(Z,N) - B(Z,N-2)],
\end{equation}
We have  calculated theoretically the results of  S$_{2n}$(Z,N) for even-even isotopes of Si, S, Ar and Ca with the help of binding energies B(Z,N) and B(Z,N-2). Comparison of theoretical results with the experimental data \cite{Wang2017} is also done in our studies.
In the Figures (\ref{Fig11} and \ref{Fig12}), we compared the theoretical  and experimental results for two neutron separation energy S$_{2n}$(Z,N)  
as a function of neutron number N for the isotopes of Silicon (left panels) and Sulphur (right panels). 
In the left panels of Figures (\ref{Fig11} and \ref{Fig12}), we observed a substantial drop in S$_{2n}$(Z,N) values at $^{28}$Si and $^{34}$Si  signaling the N = 14 and 20 shell structure in isotopes of Silicon.
In right panels of Figures (\ref{Fig11} and \ref{Fig12}), for the isotopes of Sulphur, we observed abrupt decrease in S$_{2n}$(Z,N) at $^{30}$S, $^{36}$S 
and $^{44}$S suggesting N = 14, 20 and 28 shell gaps. In Figures (\ref{Fig13} and \ref{Fig14}), we present the results for S$_{2n}$(Z,N) for isotopic chain of Argon      (left panels) and Calcium (right panels). We observed a substantial decrease in S$_{2n}$(Z,N) at $^{32}$Ar, $^{38}$Ar and $^{46}$Ar, suggesting shell gaps at N = 14, 20 and 28. However, in the case of Calcium, N = 14, 20 and 28 shell structure effects are observed at
$^{34}$Ca, $^{40}$Ca and $^{48}$Ca nuclei, respectively.
These substantial decreases in S$_{2n}$(Z,N) values in Figures (\ref{Fig11},\ref{Fig12},\ref{Fig13}   and  \ref{Fig14}) indicate that the energy required to remove two neutrons from 
(Z, N + 2) nucleus is much smaller than the energy required to remove two neutrons from (Z, N) nucleus,  where N represents the number of neutrons at 
nuclear magic numbers .

\begin{figure}
\centering
\includegraphics[trim=0 0 0 0,clip,scale=0.4]{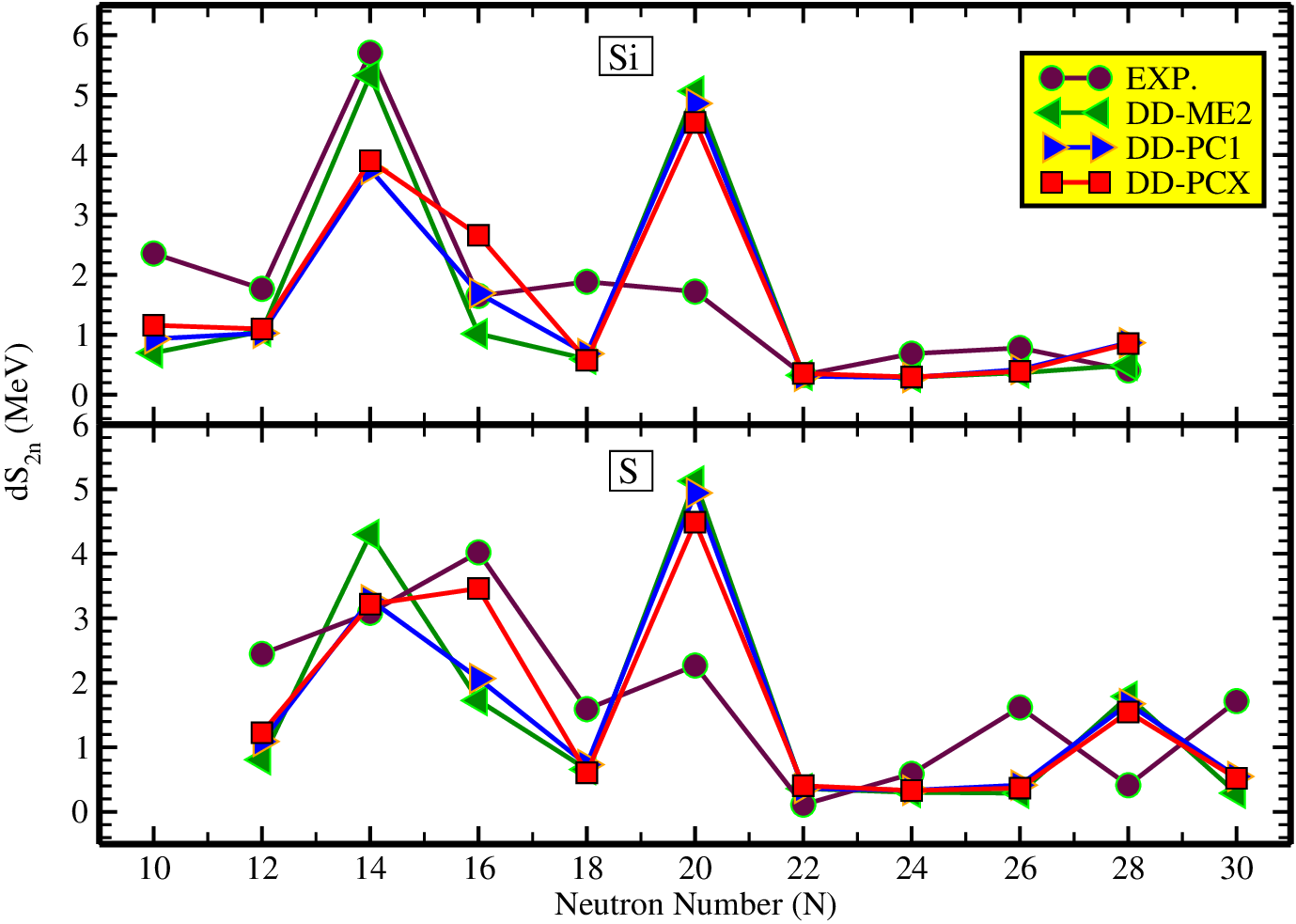}
\caption{\label{Fig15}(color online)  $dS_{2n}(Z,A)$ in units of MeV, plotted as a function of neutron number N, 
for the exotic nuclei of Silicon (left panel) and Sulphur (right panel). The theoretical estimates are computed by using Relativistic-Hartree-Bogoliubov  density functional based on Meson Exchange model parameters DD-ME2, Point Coupling model parameters DD-PC1 
and Point Coupling model parameters DD-PCX. Experimental data is taken from \cite{Wang2017}.}
\end{figure}

\begin{figure}
\centering
\includegraphics[trim=0 0 0 0,clip,scale=0.4]{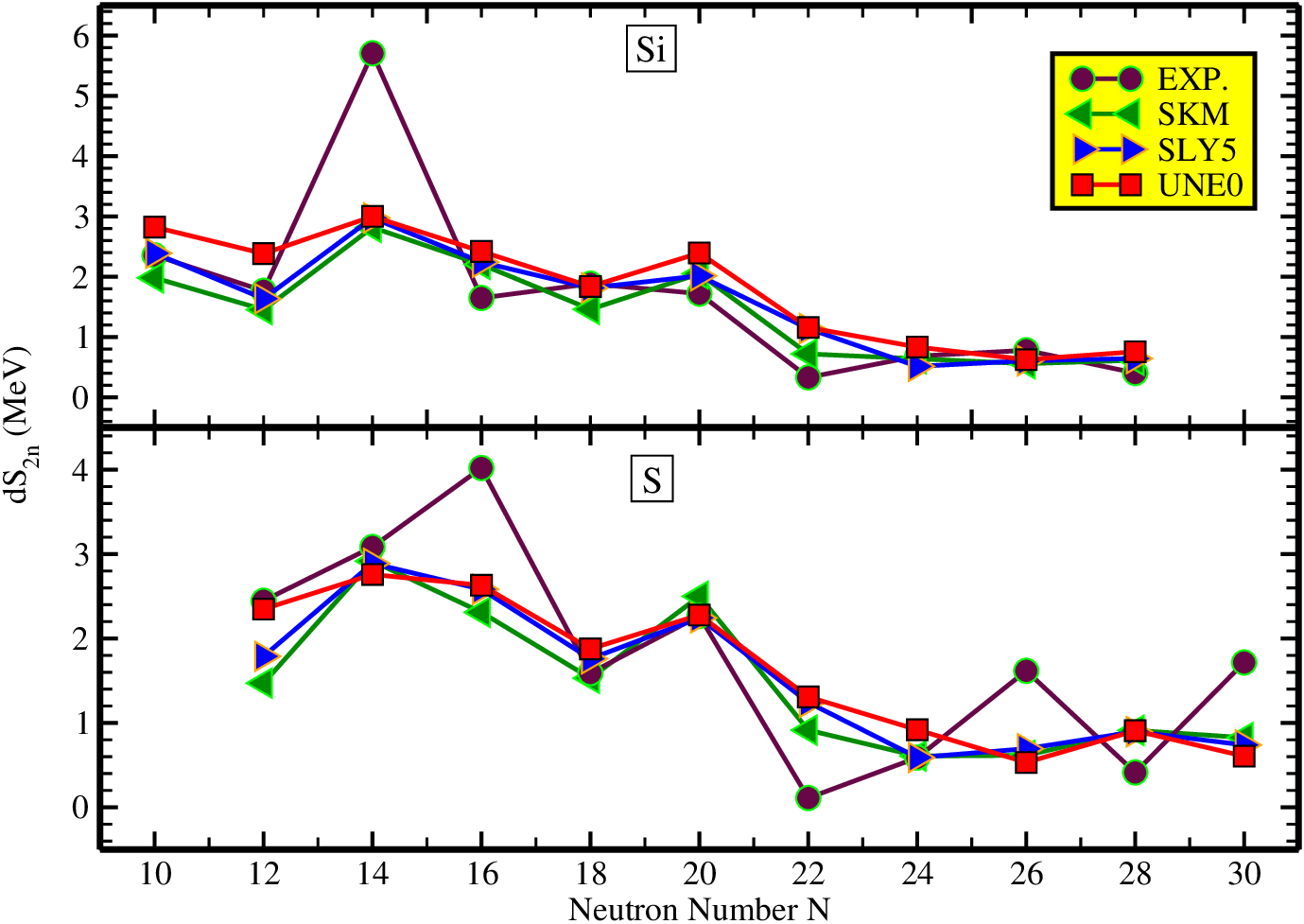}
\caption{\label{Fig16}(color online)  $dS_{2n}(Z,A)$ in units of MeV, plotted as a function of neutron number N, 
for the exotic nuclei of Silicon (left panel) and Sulphur (right panel). The theoretical estimates are computed by using Hartree-Fock-Bogoliubov  density functional based on Skyrme forces parameters SKM*,     SLY5 and UNEDF0. Experimental data is taken from \cite{Wang2017}.}
\end{figure}

\begin{figure}
\centering
\includegraphics[trim=0 0 0 0,clip,scale=0.4]{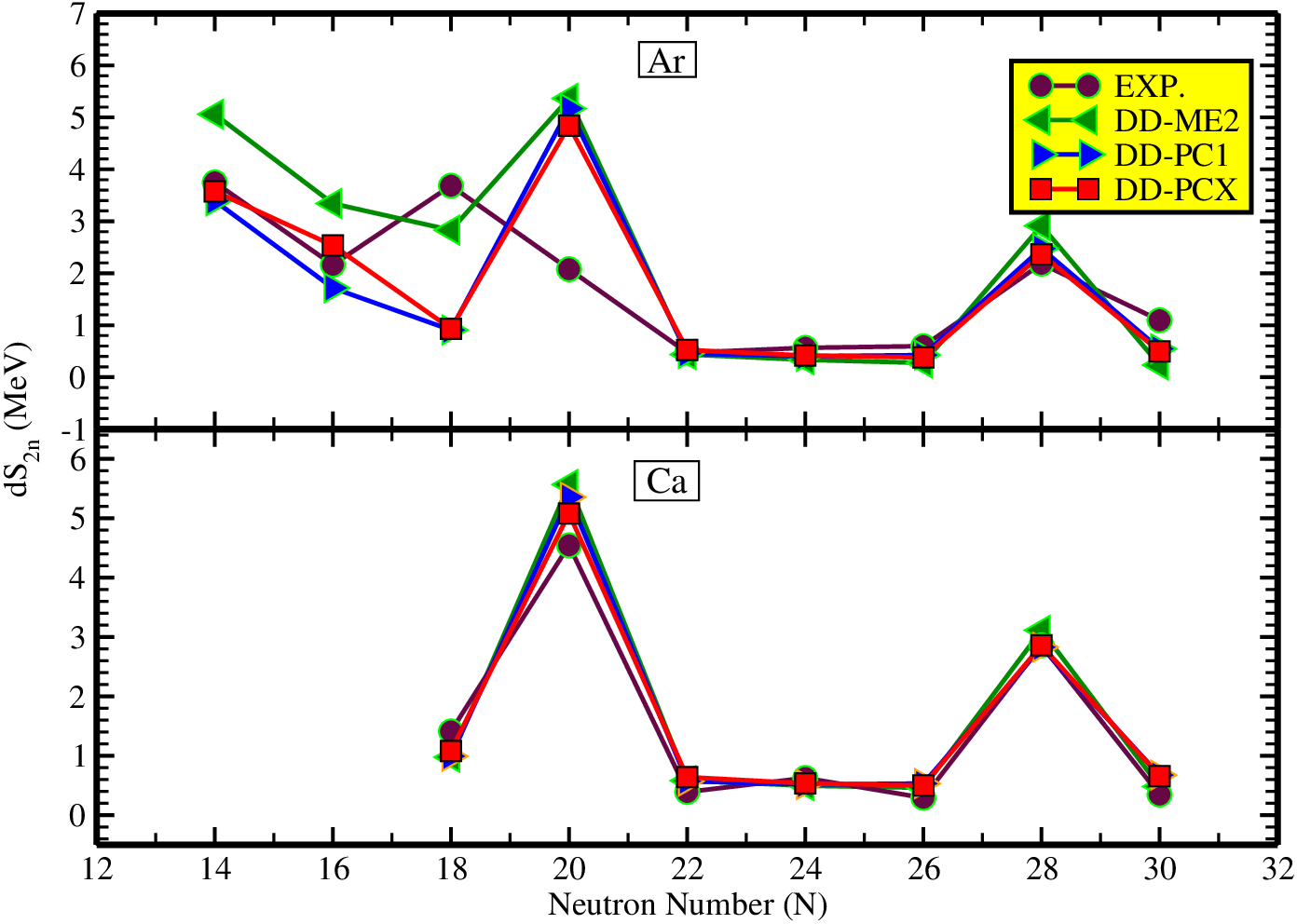}
\caption{\label{Fig17}(color online)  $dS_{2n}(Z,A)$ in units of MeV, plotted as a function of neutron number N, 
for the exotic nuclei of Argon (left panel) and Calcium (right panel). The theoretical estimates are computed by using Relativistic-Hartree-Bogoliubov  density functional based on Meson Exchange model parameters DD-ME2, Point Coupling model parameters DD-PC1 
and Point Coupling model parameters DD-PCX. Experimental data is taken from \cite{Wang2017}.}
\end{figure}

\begin{figure}
\centering
\includegraphics[trim=0 0 0 0,clip,scale=0.4]{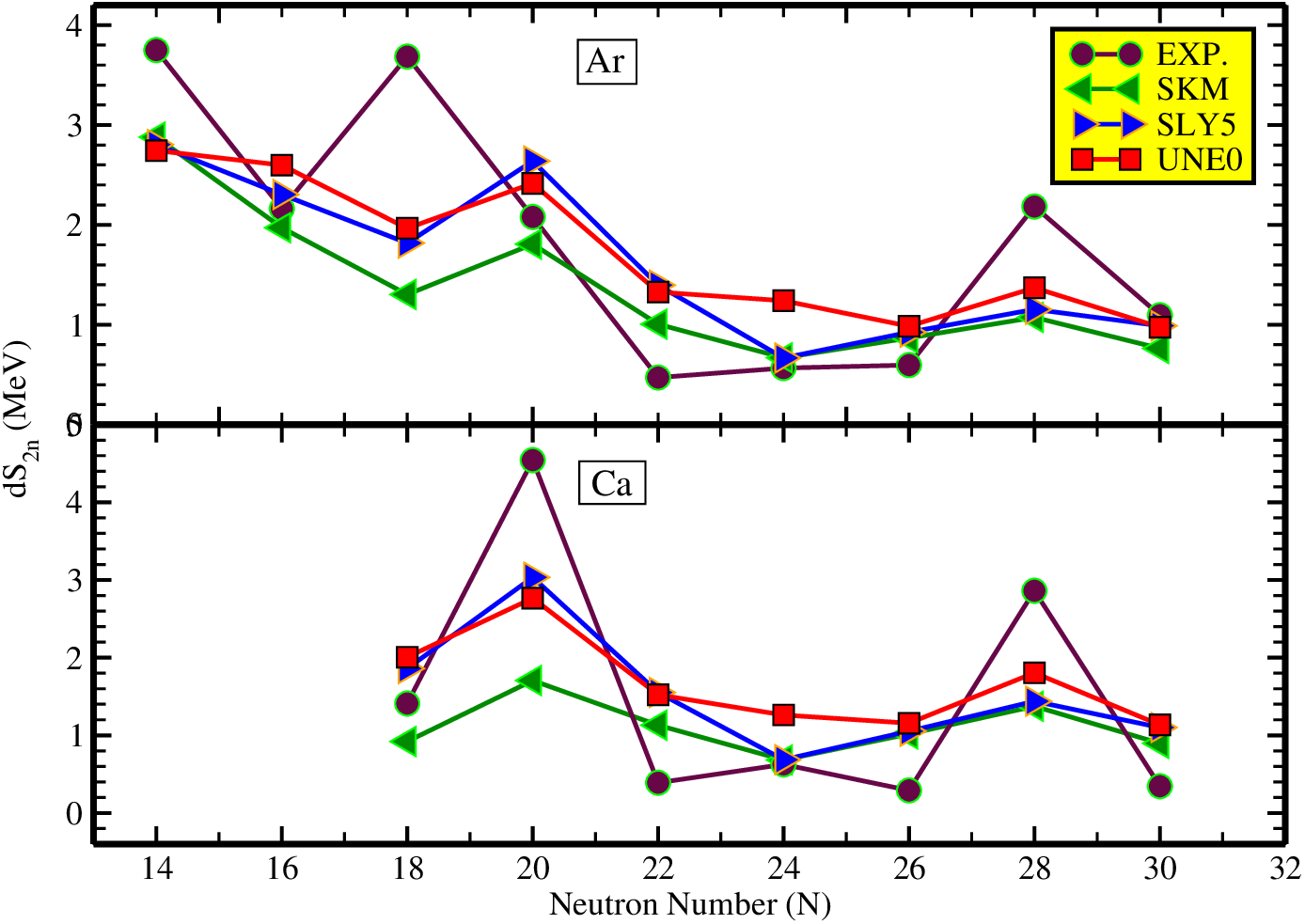}
\caption{\label{Fig18}(color online)  $dS_{2n}(Z,A)$ in units of MeV, plotted as a function of neutron number N, 
for the exotic nuclei of Argon (left panel) and Calcium (right panel). The theoretical estimates are computed by using Hartree-Fock-Bogoliubov  density functional based on Skyrme forces parameters SKM*,     SLY5 and UNEDF0. Experimental data is taken from \cite{Wang2017}.}
\end{figure}

Furthur, in order to get much deeper insight of two neutron separation energies S$_{2n}$(Z,N),  
the differential variation of the two-neutron separation energies dS$_{2n}$(Z,N) are also calculated by us using the formula  \cite{Bhuyan2015},
\begin{equation}
\label{eq:ds2n}
dS_{2n}(Z,N) = \left|\frac{S_{2n}(Z,N+2)-S_{2n}(Z,N)}{2}\right|.
\end{equation}
We have theoretically calculated dS$_{2n}$(Z,N) and then compared with experimental data of S$_{2n}$(Z,N) \cite{Wang2017}. 
In Figures (\ref{Fig15},\ref{Fig16},\ref{Fig17} and  \ref{Fig18}), we present our results of differential variation in two neutron separation energies dS$_{2n}$(Z,N) for the
even-even isotopes of Si, S, Ar and Ca, plotted against neutron number N.  It is very obvious and  intriguing to note that the trend of peaks in the $dS_{2n}(Z,N)$ values is similar to that of the discontinuities in S$_{2n}$(Z,N) values for the respective isotopic
chains of nuclides as shown in Figures (\ref{Fig11}, \ref{Fig12}, \ref{Fig13} and \ref{Fig14}).  The sharp decrease in the magnitude of dS$_{2n}$(Z,N) values on either side of a shell gap
is also unison with other physical 
observables that are already discussed in previous section for shell closures. 
 In the upper panels of Figures (\ref{Fig15} and \ref{Fig16}), for the isotopic chain of Silicon, the values of dS$_{2n}$(Z,N) have peaks at N = 14 and 20 depicting the neutron shell closures and suggesting disappearance of N = 28 magic number in $^{42}$Si nucleus which substantiate the experimental investigations reported in ref. \cite{Bastin2007}. 
 In the lower panel of Figures (\ref{Fig15} and \ref{Fig16}), we  presented the
 variation of dS$_{2n}$(Z,N) values in the isotopic chain of Sulphur, indicate peaks at N = 14 and 20. In this case, there is small peak in 
 dS$_{2n}$(Z,N) value observed in the $^{44}$S nucleus indicating shell gap at N = 28. It have been observed from experimental results \cite{42} that 
 the shell closure at N = 14 is also substantiated by the  of E$(4^+_1)$ energies, 5.17 MeV at N = 14 in $^{30}$S decreasing to E$(4^+_1) = 4.46$ MeV 
 at N = 16 in $^{32}$S nucleus. In Figures (\ref{Fig17} and  \ref{Fig18}), the peaks in the magnitude of dS$_{2n}$(Z,N) are observed at neutron number N =  20 and 28 for both isotopic chains of Argon and  Calcium.
\section{Conclusions}
We have reached to some of the striking conclusions based upon our theoretical estimations presented in this paper by employing Nuclear Density Functionals based on Relativistic-Hartree-Bogoliubov (RHB) and Hartree-Fock-Bogoliubov (HFB) theory with the different parameterizations discussed briefly in the previous sections.
The shell structure of even-even isotopes in Si, S, Ar and Ca has been extensively analysed. The
shell closure parameter $D_{n} (N)$ and the differential variation of the two-neutron separa-
tion energy $dS_{2n} (Z, N )$ are theoretically calculated and compared with the recently availabe experimental data. Our theoretical Calculations are reproducing the experimental data quite well with the tolerable discrepencies.\\
We presented the variation in binding energy per nucleon $\Delta E$ for the exotic nuclei of Si, S, Ar and Ca. The absolute value of quantity $\Delta E$ varies from 
0.01 to 0.0.26 for Si isotopes, 0.01 to 0.25 for S isotopes, 0.01 to 0.17 for Ar isotopes and 0.01 to 0.11 for the Ca isotopes as calculated with RHB theory whereas Absolute value of  $\Delta E$ varies from 0.01 to 0.22 for Si isotopes, 0.01 to 0.23 for S isotopes, 0.01 to 0.21 for Ar isotopes and 0.01 to 0.15 for the Ca isotopes when the results are calculated with HFB theory. Also, the presented theoretical extractions of shell structure properties and its interpretation with experimental data  suggests the signature of shell closure at N = 14 and
20 in case of Si, N = 14, 20 and 28 in S and N=20 and 28 in  Ar and Ca isotopes.
\ack{Virender Thakur would like to thank Himachal Pradesh University for providing computational facilities and  DST-
INSPIRE (Govt. of India) for providing financial assistance (JRF/SRF scheme) and anonymous 
refree(s) for  extremely thorough inspection of the manuscript and helpful comments.}
\section*{References}
 \bibliographystyle{iopart-num}

\begin{thebibliography}{10}
\expandafter\ifx\csname url\endcsname\relax
  \def\url#1{\texttt{#1}}\fi
\expandafter\ifx\csname urlprefix\endcsname\relax\def\urlprefix{URL }\fi
\expandafter\ifx\csname href\endcsname\relax
  \def\href#1#2{#2} \def\path#1{#1}\fi
  
\bibitem{Holt2012}
Jason~D Holt, Takaharu Otsuka, Achim Schwenk, and Toshio Suzuki.
\newblock Three-body forces and shell structure in calcium isotopes.
\newblock {\em Journal of Physics G: Nuclear and Particle Physics},
  39(8):085111, 2012.

\bibitem{Bhattacharyya2008}
S~Bhattacharyya, M~Rejmund, A~Navin, E~Caurier, F~Nowacki, A~Poves, R~Chapman,
  David O’Donnell, M~Gelin, A~Hodsdon, et~al.
\newblock Structure of neutron-rich ar isotopes beyond n= 28.
\newblock {\em Physical review letters}, 101(3):032501, 2008.
\bibitem{Lunderberg2016}
E~Lunderberg, A~Gade, V~Bader, T~Baugher, D~Bazin, JS~Berryman, BA~Brown,
  DJ~Hartley, F~Recchia, SR~Stroberg, et~al.
\newblock In-beam $\gamma$-ray spectroscopy of s 38--42.
\newblock {\em Physical Review C}, 94(6):064327, 2016.

\bibitem{Gade2009}
A~Gade, P~Adrich, D~Bazin, BA~Brown, JM~Cook, C~Aa Diget, T~Glasmacher,
  S~McDaniel, A~Ratkiewicz, K~Siwek, et~al.
\newblock In-beam $\gamma$-ray spectroscopy of very neutron-rich nuclei:
  Excited states in s 46 and ar 48.
\newblock {\em Physical review letters}, 102(18):182502, 2009.

\bibitem{Stroberg2014}
SR~Stroberg, Alexandra Gade, JA~Tostevin, VM~Bader, T~Baugher, D~Bazin,
  JS~Berryman, BA~Brown, CM~Campbell, KW~Kemper, et~al.
\newblock Single-particle structure of silicon isotopes approaching si 42.
\newblock {\em Physical Review C}, 90(3):034301, 2014.
\bibitem{Ebata2015}
Shuichiro Ebata and Masaaki Kimura.
\newblock Low-lying 2+ states generated by p n-quadrupole correlation and n= 28
  shell quenching.
\newblock {\em Physical Review C}, 91(1):014309, 2015.

\bibitem{Bastin2007}
B~Bastin, S~Gr{\'e}vy, D~Sohler, O~Sorlin, Zs~Dombr{\'a}di, NL~Achouri,
  JC~Ang{\'e}lique, F~Azaiez, D~Baiborodin, R~Borcea, et~al.
\newblock Collapse of the n= 28 shell closure in s 42 i.
\newblock {\em Physical review letters}, 99(2):022503, 2007.

\bibitem{Campbell2006}
CM~Campbell, N~Aoi, D~Bazin, MD~Bowen, B~Alex Brown, JM~Cook, D-C Dinca,
  A~Gade, T~Glasmacher, M~Horoi, et~al.
\newblock Measurement of excited states in si 40 and evidence for weakening of
  the n= 28 shell gap.
\newblock {\em Physical review letters}, 97(11):112501, 2006.

\bibitem{Takeuchi2012}
Satoshi Takeuchi, Masafumi Matsushita, Nori Aoi, P~Doornenbal, Kuoang Li, Tohru
  Motobayashi, Heiko Scheit, David Steppenbeck, He~Wang, Hidetada Baba, et~al.
\newblock Well developed deformation in si 42.
\newblock {\em Physical review letters}, 109(18):182501, 2012.
\bibitem{Force2010}
C~Force, S~Gr{\'e}vy, L~Gaudefroy, O~Sorlin, L~Caceres, F~Rotaru, J~Mrazek,
  NL~Achouri, JC~Ang{\'e}lique, F~Azaiez, et~al.
\newblock Prolate-spherical shape coexistence at n= 28 in s 44.
\newblock {\em Physical review letters}, 105(10):102501, 2010.

\bibitem{Kimura2013}
M~Kimura, Y~Taniguchi, Y~Kanada-En'yo, H~Horiuchi, and K~Ikeda.
\newblock Prolate, oblate, and triaxial shape coexistence, and the lost
  magicity of n= 28 in 43 s.
\newblock {\em Physical Review C}, 87(1):011301, 2013.

\bibitem{Sarazin2000}
F~Sarazin, H~Savajols, W~Mittig, F~Nowacki, NA~Orr, Zhongzhou Ren,
  P~Roussel-Chomaz, G~Auger, D~Baiborodin, AV~Belozyorov, et~al.
\newblock Shape coexistence and the n= 28 shell closure far from stability.
\newblock {\em Physical review letters}, 84(22):5062, 2000.

\bibitem{Gaudefroy2009}
L~Gaudefroy, JM~Daugas, M~Hass, S~Gr{\'e}vy, Ch~Stodel, JC~Thomas, L~Perrot,
  M~Girod, B~Ross{\'e}, JC~Ang{\'e}lique, et~al.
\newblock Shell erosion and shape coexistence in s 27 16 43.
\newblock {\em Physical review letters}, 102(9):092501, 2009.

\bibitem{Huck1985}
A~Huck, G~Klotz, A~Knipper, Ch~Mieh{\'e}, C~Richard-Serre, G~Walter, A~Poves,
  HL~Ravn, and G~Marguier.
\newblock Beta decay of the new isotopes k 52, ca 52, and sc 52; a test of the
  shell model far from stability.
\newblock {\em Physical Review C}, 31(6):2226, 1985.

\bibitem{Perez2017}
R~Navarro Perez, Nicolas Schunck, R-D Lasseri, C~Zhang, and Jason Sarich.
\newblock Axially deformed solution of the skyrme--hartree--fock--bogolyubov
  equations using the transformed harmonic oscillator basis (iii) hfbtho (v3.
  00): A new version of the program.
\newblock {\em Computer Physics Communications}, 220:363--375, 2017.
\bibitem{Skyrme1959}
THR Skyrme.
\newblock Thr skyrme, nucl. phys. 9, 615 (1959).
\newblock {\em Nucl. Phys.}, 9:615, 1959.

\bibitem{Nikvsic2014}
Tamara Nik{\v{s}}i{\'c}, Nils Paar, Dario Vretenar, and Peter Ring.
\newblock Dirhb—a relativistic self-consistent mean-field framework for
  atomic nuclei.
\newblock {\em Computer Physics Communications}, 185(6):1808--1821, 2014.

\bibitem{Nikvsic2010}
Tamara Nik{\v{s}}i{\'c}, Peter Ring, Dario Vretenar, Yuan Tian, and Zhong-yu
  Ma.
\newblock 3d relativistic hartree-bogoliubov model with a separable pairing
  interaction: Triaxial ground-state shapes.
\newblock {\em Physical Review C}, 81(5):054318, 2010.

\bibitem{Ring1980}
Peter Ring and Peter Schuck.
\newblock {\em The nuclear many-body problem}.
\newblock Springer Science \& Business Media, 2004.

\bibitem{Stoitsov2013}
MV~Stoitsov, Nicolas Schunck, Markus Kortelainen, N~Michel, H~Nam, E~Olsen,
  Jason Sarich, and S~Wild.
\newblock Axially deformed solution of the skyrme-hartree--fock--bogoliubov
  equations using the transformed harmonic oscillator basis (ii) hfbtho v2.
  00d: A new version of the program.
\newblock {\em Computer Physics Communications}, 184(6):1592--1604, 2013.

\bibitem{Nikvsic2011}
T~Nik{\v{s}}i{\'c}, Dario Vretenar, and Peter Ring.
\newblock Relativistic nuclear energy density functionals: Mean-field and
  beyond.
\newblock {\em Progress in Particle and Nuclear Physics}, 66(3):519--548, 2011.

\bibitem{Thakur2019}
Virender Thakur and Shashi~K Dhiman.
\newblock A study of charge radii and neutron skin thickness near nuclear drip
  lines.
\newblock {\em Nuclear Physics A}, 2019.

\bibitem{Chabanat1997}
E~Chabanat, P~Bonche, P~Haensel, J~Meyer, and R~Schaeffer.
\newblock A skyrme parametrization from subnuclear to neutron star densities.
\newblock {\em Nuclear Physics A}, 627(4):710--746, 1997.

\bibitem{Kortelainen2010}
Markus Kortelainen, Thomas Lesinski, J~Mor{\'e}, W~Nazarewicz, J~Sarich,
  N~Schunck, MV~Stoitsov, and S~Wild.
\newblock Nuclear energy density optimization.
\newblock {\em Physical Review C}, 82(2):024313, 2010.

\bibitem{Kortelainen2012}
M~Kortelainen, J~McDonnell, Witold Nazarewicz, P-G Reinhard, J~Sarich,
  N~Schunck, MV~Stoitsov, and SM~Wild.
\newblock Nuclear energy density optimization: Large deformations.
\newblock {\em Physical Review C}, 85(2):024304, 2012.

\bibitem{Bartel1982}
J~Bartel, Ph~Quentin, Matthias Brack, C~Guet, and H-B H{\aa}kansson.
\newblock Towards a better parametrisation of skyrme-like effective forces: A
  critical study of the skm force.
\newblock {\em Nuclear Physics A}, 386(1):79--100, 1982.

\bibitem{Dobaczewski1984}
J~Dobaczewski, H~Flocard, and J~Treiner.
\newblock Hartree-fock-bogolyubov description of nuclei near the neutron-drip
  line.
\newblock {\em Nuclear Physics A}, 422(1):103--139, 1984.

\bibitem{Stoitsov2005}
MV~Stoitsov, J~Dobaczewski, Witold Nazarewicz, and P~Ring.
\newblock Axially deformed solution of the skyrme--hartree--fock--bogolyubov
  equations using the transformed harmonic oscillator basis. the program hfbtho
  (v1. 66p).
\newblock {\em Computer physics communications}, 167(1):43--63, 2005.

\bibitem{Lalazissis2005}
GA~Lalazissis, Tamara Nik{\v{s}}i{\'c}, Dario Vretenar, and Peter Ring.
\newblock New relativistic mean-field interaction with density-dependent
  meson-nucleon couplings.
\newblock {\em Physical Review C}, 71(2):024312, 2005.

\bibitem{Nikvsic2008}
Tamara Nik{\v{s}}i{\'c}, Dario Vretenar, and Peter Ring.
\newblock Relativistic nuclear energy density functionals: Adjusting parameters
  to binding energies.
\newblock {\em Physical Review C}, 78(3):034318, 2008.
\bibitem{Vretenar2005}
Dario Vretenar, AV~Afanasjev, GA~Lalazissis, and P~Ring.
\newblock Relativistic hartree--bogoliubov theory: static and dynamic aspects
  of exotic nuclear structure.
\newblock {\em Physics reports}, 409(3-4):101--259, 2005.

\bibitem{Meng2006}
Jie Meng, H~Toki, Shan-Gui Zhou, SQ~Zhang, WH~Long, and LS~Geng.
\newblock Relativistic continuum hartree bogoliubov theory for ground-state
  properties of exotic nuclei.
\newblock {\em Progress in Particle and Nuclear Physics}, 57(2):470--563, 2006.

\bibitem{Ring1997}
P~Ring, YK~Gambhir, and GA~Lalazissis.
\newblock Computer program for the relativistic mean field description of the
  ground state properties of even-even axially deformed nuclei.
\newblock {\em Computer physics communications}, 105(1):77--97, 1997.

\bibitem{Gonzalez1996}
T~Gonzalez-Llarena, JL~Egido, GA~Lalazissis, and P~Ring.
\newblock Relativistic hartree-bogoliubov calculations with finite range
  pairing forces.
\newblock {\em Physics Letters B}, 379(1-4):13--19, 1996.
\bibitem{Tian2009}
Yuan Tian, ZY~Ma, and P~Ring.
\newblock A finite range pairing force for density functional theory in
  superfluid nuclei.
\newblock {\em Physics Letters B}, 676(1-3):44--50, 2009.

\bibitem{Yuksel2019}
E~Y{\"u}ksel, T~Marketin, and N~Paar.
\newblock Optimizing the relativistic energy density functional with nuclear
  ground state and collective excitation properties.
\newblock {\em Physical Review C}, 99(3):034318, 2019.

\bibitem{Wang2017}
Meng Wang, G~Audi, FG~Kondev, WJ~Huang, S~Naimi, and Xing Xu.
\newblock The ame2016 atomic mass evaluation (ii). tables, graphs and
  references.
\newblock {\em Chinese Physics C}, 41(3):030003, 2017.

\bibitem{Brown2013}
B.~Alex Brown.
\newblock Nuclear pairing gap: How low can it go?
\newblock {\em Phys. Rev. Lett.}, 111:162502, Oct 2013.
\bibitem{Thakur2019study}
Smriti Thakur and Shashi~K Dhiman.
\newblock A study of shell structure in normal to exotic nuclides within
  relativistic hartree--bogoliubov approximation.
\newblock {\em Modern Physics Letters A}, 34(02):1950014, 2019.
\bibitem{Bhuyan2015}
M~Bhuyan.
\newblock Structural evolution in transitional nuclei of mass 82≤ a≤ 132.
\newblock {\em Physical Review C}, 92(3):034323, 2015.

\bibitem{Crawford2010}
HL~Crawford, RVF Janssens, PF~Mantica, JS~Berryman, R~Broda, MP~Carpenter,
  N~Cieplicka, B~Fornal, GF~Grinyer, N~Hoteling, et~al.
\newblock $\beta$ decay and isomeric properties of neutron-rich ca and sc
  isotopes.
\newblock {\em Physical Review C}, 82(1):014311, 2010.

\bibitem{Rosenbusch2015}
M~Rosenbusch, Pauline Ascher, Dinko Atanasov, C~Barbieri, D~Beck, Klaus Blaum,
  Ch~Borgmann, M~Breitenfeldt, R~Burcu Cakirli, A~Cipollone, et~al.
\newblock Probing the n= 32 shell closure below the magic proton number z= 20:
  Mass measurements of the exotic isotopes k 52, 53.
\newblock {\em Physical review letters}, 114(20):202501, 2015.

\bibitem{42}
{\em http://www.nndc.bnl.gov}.

\bibitem{Grasso2014}
Marcella Grasso.
\newblock Magicity of the ca 52 and ca 54 isotopes and tensor contribution
  within a mean-field approach.
\newblock {\em Physical Review C}, 89(3):034316, 2014.
\bibitem{Chasman1976}
R.~R. Chasman.
\newblock Density-dependent delta interactions and actinide pairing matrix
  elements.
\newblock {\em Phys. Rev. C}, 14:1935--1945, Nov 1976.
\bibitem{Terasaki1997}
Jun Terasaki, Heenen Flocard, P-H Heenen, and Paul Bonche.
\newblock Deformation of nuclei close to the two-neutron drip line in the mg
  region.
\newblock {\em Nuclear Physics A}, 621(3):706--718, 1997.
\bibitem{Dobaczewski2002}
J~Dobaczewski, W~Nazarewicz, and MV~Stoitsov.
\newblock Nuclear ground-state properties from mean-field calculations.
\newblock pages 55--60, 2003.
  
\end{thebibliography}

\Figures
\end{document}